\documentclass[twocolumn,notitlepage,aps,pra,superscriptaddress,showpacs,floatfix,longbibliography]{revtex4-1}
\usepackage{graphics,epsfig,graphicx}
\usepackage{amsbsy}
\usepackage{amsmath}
\usepackage{bm}
\usepackage{amsfonts}
\usepackage{cancel}
\usepackage{multirow}
\usepackage{cancel}
\begin{document}

\title{Embedding nuclear physics inside the unitary window}

\author{M. Gattobigio}
\affiliation{
 Universit\'e C\^ote d'Azur, CNRS, Institut  de  Physique  de  Nice,
1361 route des Lucioles, 06560 Valbonne, France }
\author{A. Kievsky} 
\author{M. Viviani}
\affiliation{Istituto Nazionale di Fisica Nucleare, Largo Pontecorvo 3, 56100 Pisa, Italy}
\begin{abstract}
  The large values of the singlet and triplet scattering lengths locate the
  two-nucleon system close to the unitary limit, the limit in which these two
  values diverge. As a consequence, the system shows a continuous scale
  invariance which strongly constrains the values of the observables, a
  well-known fact already noticed a long time ago. The three-nucleon system
  shows a discrete scale invariance that can be observed by correlations of the
  triton binding energy with other observables as the doublet nucleon-deuteron
  scattering length or the alpha-particle binding energy. The low-energy
  dynamics of these systems is universal; it does not depend on the details of
  the particular way in which the nucleons interact. Instead, it depends on a
  few control parameters, the large values of the scattering lengths and the
  triton binding energy. Using a potential model with variable strength set to
  give values to the control parameters, we study the spectrum of $A=2,3,4,6$
  nuclei in the region between the unitary limit and their physical values. In
  particular, we analyze how the binding energies emerge from the unitary limit
  forming the observed levels.
\end{abstract}
\maketitle

\section{Introduction} 

  In a non-relativistic theory where one allows for a
  tunable potential (or other parameters like  the mass $m$ of the particles), we
  refer to the unitary window as to the range of those parameters for which the
  scattering(s) length(s) $a$ attains a value close to infinity (the unitary
  limit). 
  This is a relevant limit because the physics becomes
  universal~\cite{braaten:2006_PhysicsReports} and a common description can be
  used for totally different systems, ranging from nuclear physics up to atomic
  physics or down in scale to hadronic systems. For instance, in the two-body
  sector there is the appearance of a shallow (real or virtual) bound state
  whose energy is governed by the scattering length, $E_2=-\hbar^2/ma^2$.
  This state is shallow if compared with the energy related to the typical
  interaction length $\ell$, defined as $-\hbar^2/m\ell^2$, and in the limit
  $a/\ell\rightarrow\infty$,  where it becomes resonant. This limit can be understood  
  either as the scattering length going to infinity or as the range of the
  interaction going to zero; in the last case one talks of zero-range limit
  or scaling limit. 
  
  In the scaling limit, the two-body sector displays continuous scale invariance
  due to the fact that the only dimensionful parameter is the scattering length.
  As soon as we change the number of particles, the above symmetry is
  dynamically broken to a discrete scale invariance (DSI); for example, for three equal
  bosons at the unitary limit, an infinite tower of bound states appears
  - the Efimov effect~\cite{efimov:1970_Phys.Lett.B,efimov:1971_Sov.J.Nucl.Phys.} -
  related by a discrete scale transformation $r\rightarrow \exp(\pi/s_0) \,r
  \approx 22.7\,r$, with the scaling factor $s_0=1.0062\dots$ a universal
  transcendental number that does not depend on the actual physical
  system.  The
  anomalous breaking of the symmetry gives rise to an emergent scale at the
  three-body level which is usually referred to as the three-body parameter
  $\kappa_*$, giving the binding energy $\hbar^2\kappa_*^2/m$ of a reference state
  of the above tower of states.

  The fine tuning of the parameters that brings a systems inside the
  unitary window can be realized either artificially, like it
  has been the case in the field of cold atoms with Feshbach
  resonances~\cite{chin:2010_Rev.Mod.Phys.}, or can be provided by nature, as in
  the case of atomic $^4$He, where the $^4$He$_2$ molecule has a binding energy
  of several order smaller than the typical interaction
  energy~\cite{luo:1993_J.Chem.Phys.}. Nuclear physics is another example of 
  tuned-by-nature system; the binding energy of  the  
  deuteron, $B_d=2.22456$~MeV is much smaller then the typical-nuclear
  energy $\hbar^2/m\ell^2 \approx 20$~MeV , considering that the 
  interaction length is given by the inverse of the pion mass $m_\pi$, 
  $\ell\sim 1/m_\pi \approx 1.4$~fm. The fact that nuclear physics resides
  inside such a window has been used in the pioneering works of the thirties
  where the binding energies of light nuclei have been calculated using either boundary
  conditions~\cite{wigner:1933_Phys.Rev.,fermi:1936_Ricercasci.} or
  pseudopotentials~\cite{huang:1957_Phys.Rev.}.

  Nuclear physics is the low energy aspect of the strong interaction, namely
  Quantum Chromo Dynamics (QCD); in this limit, QCD is a strong interacting
  quantum field theory and only non-perturbative approaches could be used to
  describe the spectrum of nuclear physics. Such non-perturbative approaches
  start to appear, one example being Lattice QCD (LQCD); however, a complete calculation 
  of nuclear properties seems at present not yet feasible using these techniques.
  Historically, the description of light nuclear systems were based on potential
  models constructed to reproduce a selected number of observables; first
  attempts have been based on the expansion of the potential on the most general
  operator basis compatible with the symmetries observed in the spectrum. 
  Lately, it has been realized that a potential could be constructed starting
  from the symmetries of QCD in an Effective Field Theory (EFT) approach. One of
  the important symmetry of QCD in the limit of zero-mass light quarks is the
  Chiral Symmetry; this symmetry is indeed spontaneous broken and its
  Goldstone boson is the $\pi$-meson. The mass of the pion $m_\pi$ is not really zero
  because of the soft breaking term introduced by the explicit mass of the 
  quarks up and down, but still is much lower than the typical hadronic masses.
  The chiral limit is not the only interesting limit in QCD; the actual mass of
  the pion is probably close to a value for which the nucleon-nucleon scattering
  lengths diverge~\cite{braaten:2003_Phys.Rev.Lett.}; in fact, one can study the
  variation of the $^1S_0$ (singlet) $a_0$ and $^3S_1$ (triplet) $a_1$
  scattering lengths as a function of the masses of the up and down quarks, or
  equivalently of $m_\pi$ which is related to the quark mass by the
  Gell-Mann-Oakes-Renner relation. It has been shown that for $m_\pi\approx
  200$~MeV both scattering lengths go to
  infinity~\cite{epelbaum:2006_Eur.Phys.J.C,beane:2002_Nucl.Phys.A} . At the
  physical point, $m_\pi\approx138$~MeV, the values of the two scattering
  lengths, $a_0 \approx -23.7$~fm and $a_1\approx 5.4$~fm, are still (much) larger than
  the typical interaction length $\ell\approx 1.4$~fm; this is a further
  indication that nuclear physics is close to the unitary limit and well
  inside the universal window.

A model-independent description of the physics inside the unitary window  is
given by an EFT based on the clear separation of scales between the typical
momenta~$Q \sim 1/a$ of the system and the underling high momentum scale $\sim
1/\ell$~\cite{vankolck:1999_NuclearPhysicsA,bedaque:1999_Phys.Rev.Lett.,bedaque:1999_Nucl.Phys.A}.
Using EFT, if the power-counting is
correct~\cite{griesshammer:2005_Nucl.Phys.A}, one can systematically improve the
prediction on observables. For instance, in the two-body sector the usual
effective range expansion (ERE) can be reproduced by such an
expansion~\cite{vankolck:1999_NuclearPhysicsA}; the leading order (LO), which is
just a two-body contact interaction, captures all the information encoded in the
scattering length $a$, while the next-to-the-leading order term (NLO), which
contains derivatives, captures the information encoded in the effective range
$r_e$. Starting from the three-body sector, a LO three-body interaction is
necessary~\cite{bedaque:1999_Phys.Rev.Lett.,bedaque:1999_Nucl.Phys.A,kievsky:2017_Phys.Rev.C} which
introduces the emergent three-body scale.

It is possible to investigate the universal window by using potential
models; this approach allows to follow the behaviour of two- and
three-particle binding energies inside the window of universality. Also 
a higher number of particles can be considered as in 
Refs.~\cite{gattobigio:2011_Phys.Rev.A,kievsky:2014_Phys.Rev.A,kievsky:2017_Phys.Rev.A} where it
has been shown that the use of a simple Gaussian potential gives a good description
of bosonic systems like  Helium droplets in this regime. 

In this paper we want to explore the window of universality for nucleons, that
means for fermions with 1/2 spin and isospin degrees of freedom; the idea is to
follow, as a function of the interaction range, the states which represent light
nuclei in the region of universality and to observe which part of the nuclear
spectrum is in fact governed by universality.
The major difference with respect the bosonic case is the presence of two
scattering lengths. There has been previous studies of the Efimov physics with
two scattering
lengths~\cite{bulgac:1976_Sov.J.Nucl.Phys.,kievsky:2016_Few-BodySyst,%
kievsky:2018_J.Phys.Conf.Ser.,konig:2017_Phys.Rev.Lett.,%
vankolck:2018_J.Phys.Conf.Ser.,hammer:2018_Few-BodySyst.,beane:2002_Nucl.Phys.A},
and there are different ways to explore the space of parameters; one possible choice
is to keep constant the ration between the scattering lengths $a_0/a_1$,
selecting some cuts for in that space. Accordingly, we explore the nuclear cut, 
that means $a_0/a_1\approx -4.3$, moving from the unitary point,
$a_0,a_1\rightarrow\infty$, to the physical point; at a more fundamental
level, this is equivalent to change the mass of the pion $m_\pi$ ( or the sum of
up and quark masses in QCD), as it was shown in
Refs.~\cite{beane:2001_Phys.Rev.A,epelbaum:2006_Eur.Phys.J.C}.  Interestingly,
we observe that, at unitary, in addition to the $A=5$ gap we observe   
a  $A=6$ gap. 

The paper is organized at follows. In
Section~\ref{sec:efimovPlot} we will show how the spectrum of $A=2,3,4,6$
nucleons representing light nuclei depend on the scattering lengths when
we change them from the unitary limit to the their real value, and we
discuss what are the aspects of the universality of Efimov physics that still
remain. In Section~\ref{sec:physicalPoint} we concentrate our study at the 
physical point, where a three-body force, as well as the Coulomb interaction,
are introduced. In Section~\ref{sec:pwaves} we investigate the possible r\^ole of 
$p$-waves in the binding of $A=6$ nuclei. Finally, in
Section~\ref{sec:conclusions} we give our conclusions.

\section{$1/2$ Spin-Isospin energy levels close to unitary}\label{sec:efimovPlot}
In this section, we describe the discrete spectrum of spin $1/2$-
isospin $1/2$ particles from the unitary limit to the point where
nuclear physics is located, the physical point, defined as the point in which
the scattering lengths take their observed values. To this end we construct
the Efimov plot, a plot in the plane $(K,1/a)$ defined by the energy
momentum $K$ of the bound state with energy $\hbar^2K^2/m$, as a function of the inverse of
the two-body scattering length $a$. In the case of two nucleons there are
two different scattering lengths, $a_0$ and $a_1$, in
spin-isospin channels with $S,T=0,1$ and $S,T=1,0$ respectively. Accordingly, following
Refs.~\cite{kievsky:2016_Few-BodySyst,kievsky:2018_J.Phys.Conf.Ser.}, the plane
is defined with the triplet scattering length $(K,1/a_1)$, taking care that for
each value of $a_1$, $a_0$ is varied accordingly maintaining the ratio $a_0/a_1$ constant.  
In Refs.~\cite{kievsky:2016_Few-BodySyst,kievsky:2018_J.Phys.Conf.Ser.}, 
the main characteristic of the Efimov plot for three $1/2$
spin-isospin particles have been studied. In particular it was shown that for the ratio
$a_0/a_1\approx-4.3$, corresponding to the nuclear physics case, the infinite
tower of states at unitary disappear very fast as $a_1$ decreases and, at
$a_1<20\;$fm, only one state survives. This simple analysis explains the
existence of only one bound state for $^3$H and $^3$He. Conversely, in the case
of three identical bosons, calculations using finite-range potentials have shown
that the first excited state survive along the unitary window.

\subsection{The potential model}

In order to explore the unitary window through the Efimov plot, we calculate the
binding energies of $A$ nucleons for different values of the two-body scattering
lengths. In the case of a zero-range interaction the $A=2$ energy of the real
(virtual) state for $a>0$ ($a<0$) is simply $E_2=-\hbar^2/m a^2$. For three
particles, and using a zero-range interaction, the binding energies can be
obtained by solving the Faddeev zero-range equations encoded in the
Skorniakov-Tern-Martirosian equations (see
Ref.~\cite{braaten:2006_PhysicsReports} and references there in for details). It
is well known that the contact interaction can be represented by different
functional forms introducing finite-range effects. In particular, as it has been
shown in Ref.~\cite{alvarez-rodriguez:2016_Phys.Rev.A}, inside the unitary
window a Gaussian potential captures the main characteristics of the dynamics,
confirming the universal behavior of the system in this particular region.
Considering that, for two nucleons, there are four different spin-isospin
channels with quantum numbers $ST=01,10,00,11$, we define the following
spin-isospin dependent potential of a Gaussian type
\begin{equation}
  V(r)  = \sum_{ST} V_{ST}\,e^{-(r/r_{ST})^2} {\cal
  P}_{ST} \,,
  \label{eq:twoBody}
\end{equation}
where we have introduced the spin-isospin $ {\cal P}_{ST}$ projectors. 
The minimal requirement to construct a fully antisymmetric two-body wave function
with the lowest value of the angular momentum $L$ is to consider
the spin-isospin channels $S=0,T=1$ and $S=1,T=0$. 
Therefore, in this first analysis, the other two components of the potential are set
to zero: $V_{00}=V_{11}=0$.
In each of the two remaining terms there are two parameters, the strength of the
Gaussian and its range; we fix both ranges to be the same $r_{10} = r_{01} =
r_0 = 1.65$~fm, and of the order of the nuclear range. With this choice
an acceptable description of the two-body low-energy data is obtained 
(a refinement of the model will be discussed in the next section).
The tuning of the two strengths allows us to
control the scattering lengths; the value of $V_{01}$ defines the singlet
scattering length $a_0$, while the value of $V_{10}$ defines triplet one $a_1$. 
In all our calculations we fix the value of the nucleon mass $m$ so that
$\hbar^2/m=41.47$~MeV~$\text{fm}^2$. 
In some of the following Table/Figures, as unit length we use we use
$r_0=1.65$~fm and as unit of energy we use $E_0 = \hbar^2/mr_0^2 = 15.232$~MeV.

In order to calculate the binding energies for the nuclear systems having
$A=3,4,6$, we have solved the Schr\"odinger equation using two different variational methods. One method
is based on the Hyperspherical Harmonic (HH)~\cite{kievsky:1997_Few-BodySyst}
basis in its unsymmetrized
version~\cite{gattobigio:2009_Phys.Rev.A,gattobigio:2009_Few-BodySyst.,gattobigio:2011_Phys.Rev.C}.
We have used this approach mainly for $A=3,4$ since it is very accurate for states far from thresholds. 
Close to a threshold, as for $A=6$ or for excited stated in $A=3,4$, the dimension of the basis tends 
to become too big to have a good precision. To overcome this problem we implemented a version of the 
stochastic variational method (SVM)~\cite{varga:1995_Phys.Rev.C} with correlated-Gaussian functions as
basis set; this method allows for a more economical description of
the excited states close to the threshold. 

By giving values to $V_{10}$ and $V_{01}$, the values of the scattering lengths
vary along the nuclear cut defined from the ratio $a_0/a_1 = -4.3066$. Along
this path we have calculated the binding energies of $A=2,3,4,6$ nucleon systems. 
The calculations, for selected values of the potential strengths,
 are reported in Table~\ref{tab:nuclearPlane} in the case of positive
triplet-scattering length values, for which a two-body bound state in the
$^3S_1$ channel exists. The calculations cover a region
between the unitary point, for which both scattering lengths attain an infinite value, and the
physical point, for which the value of the two-body state is $E_2 = -2.2255$~MeV
(the experimental binding energy of the deuteron is $2.224575(9)$~MeV), and the two
scattering lengths have the values $a_1 = 5.4802$~fm and $a_0=-23.601$~fm, with
the experimental values $a_1=5.424(3)$~fm and $a_0=-23.74(2)$~fm, respectively.

\begin{table*}[ht]
  \caption{Calculations belonging to the nuclear cut, $a_0/a_1 =
    -4.3066 $ for selected values of the strengths $V_{01}$ and $V_{10}$.
   The ground-state energy $E_A$ and, if it exists, the excited-state energy 
   $E_A^*$ of the $A$-particle system are reported. In the $A=6$ case we distinguish 
   between the total isospin $T=1$ and total spin $S=0$ case, $^6$He, and the 
   $T=0$, $S=1$ case, $^6$Li. The Coulomb interaction is not taken 
  into account.}
  \label{tab:nuclearPlane}
\begin{tabular} {@{}l l c c c c c c c c@{}}
\hline\hline
$V_{10}$(MeV)& $V_{01}$(MeV)  & $a_1$(fm) & $E_2$(MeV) & $E_3$(MeV) &
$E_3^*$(MeV) & $E_4$(MeV) & $E_4^*$(MeV) & $^6$He(MeV) &  $^6$Li(MeV)\\ 
\hline 
-60.575  &  -37.9          &  5.4802      &  -2.2255       &  -10.2455  &  -           &   -39.843   &  -11.19  & -41.60 & -46.74 \\
-60.     &  -37.95858685   &  5.5980      &  -2.1098       &  -10.0056  &  -           &   -39.221   &  -10.93  & -40.87 & -45.82 \\
-58.     &  -38.17113668   &  6.0683      &  -1.72703      &  -9.19027  &  -           &   -37.093   &  -10.01  & -38.36 & -42.71 \\
-56.     &  -38.39860618   &  6.6607      &  -1.37621      &  -8.40544  &  -           &   -35.017   &   -9.14  & -35.95 & -39.67 \\
-54.     &  -38.64295075   &  7.4310      &  -1.05929      &  -7.65258  &  -           &   -32.997   &   -8.31  & -33.58 & -36.77 \\
-52.     &  -38.90658498   &  8.4756      &  -0.77842      &  -6.93330  &  -           &   -31.035   &   -7.52  & -31.31 & -33.95 \\
-50.0    &  -39.19224      &  9.97497     &  -0.53599      &  -6.24929  &  -           &   -29.135   &   -6.78  &  -     & -31.23 \\
-48.0    &  -39.50320907   &  12.31255    &  -0.334659     &  -5.60235  &  -           &   -27.300   &   -6.08  &  -     & -28.62 \\
-46.0    &  -39.8434712    &  16.47151    &  -0.17735880   &  -4.99446  &  -           &   -25.536   &   -5.43  &  -     & -26.17  \\ 
-45.0    &  -40.026055     &  20.06376    &  -0.1163       &  -4.7058   & -0.116853    &   -24.682   &   -5.13  &  -     & -24.96  \\ 
-44.5    &  -40.120751     &  22.6040702  &  -0.0903760    &  -4.5654   & -0.091991    &   -24.262   &   -4.98  &  -     & -24.41  \\ 
-44.0    &  -40.217947     &  25.95893    &  -0.067559     &  -4.4278   & -0.07054     &   -23.847   &   -4.83  &  -     &   -     \\
-43.5    &  -40.3174375    &  30.5953     &  -0.047939     &  -4.2927   & -0.053034    &   -23.437   &   -4.69  &  -     &   -     \\
-43.0    &  -40.4196253    &  37.421571   &  -0.0315796    &  -4.1605   & -0.03853     &   -23.032   &   -4.55  &  -     &   -     \\
-42.5    &  -40.52452499   &  48.46985    &  -0.0185467    &  -4.0311   & -0.02697     &   -22.633   &   -4.42  &  -     &   -      \\
-42.0    &  -40.63225234   &  69.413066   &  -0.0089083    &  -3.9044   & -0.01816     &   -22.238   &   -4.28  &  -     &   -     \\
-41.5    &  -40.74293099   &  124.3314    &  -0.00273453   &  -3.7807   & -0.01186     &   -21.850   &   -4.15  &  -     &   -     \\
-40.88363&  -40.88363      &  $\infty$    &  0             &  -3.6322   & -0.00678     &   -21.378   &   -4.00  &  -     &   -     \\
 \hline
\hline
\end{tabular}
\end{table*}
Considering the potential of Eq.(\ref{eq:twoBody}) in all cases the lowest state corresponds
to total orbital angular momentum $L=0$. Moreover, in this first analysis the Coulomb interaction 
between protons were disregarded, so the isospin is conserved. 
In the three-body sector, the quantum numbers of 
$^3$H and of $^3$He are $S=1/2$ and $T=1/2$. In this exploration we 
disregarded other charge-symmetry breaking terms, accordingly
the two nuclei have the same energy. We refer to their ground-state
energy as $E_3$ and to their excited-state energy as $E_3^*$. The 
total wave function is antisymmetric with the spatial wave function 
mostly symmetric. We would like to stress a big difference between the bosonic and the nuclear cut
already mentioned above: in the bosonic case the first
excited state never disappears into the particle-dimer continuum whereas,
in the nuclear case, the excited state disappear in the continuum and 
it becomes a virtual state already at a large value of $a_1$ ($a_1\approx
20$~fm). The motivation
is the following: at unitary, since we are using the same range for both
gaussians, the system is equivalent to a bosonic system and an infinite set
of excited states appears showing the Efimov effect (in
Table~\ref{tab:nuclearPlane} only the first
one is reported). Moreover, the system is completely symmetric, no other
symmetry is present. As the strength of the potentials starts to vary, keeping
the ratio $a_0/a_1$ constant,
the three-body wave function develops a spatial mixed symmetry component making
the energy gain slower than in the bosonic case. The two-body system is not affected 
by the singlet potential (which is smaller) and its energy gain is the same as
in the bosonic case; as a consequence the first excited state
crosses the particle-dimer continuum becoming a virtual state.

\begin{figure}[!tbp]
  \begin{center} 
    \includegraphics[width=\linewidth]{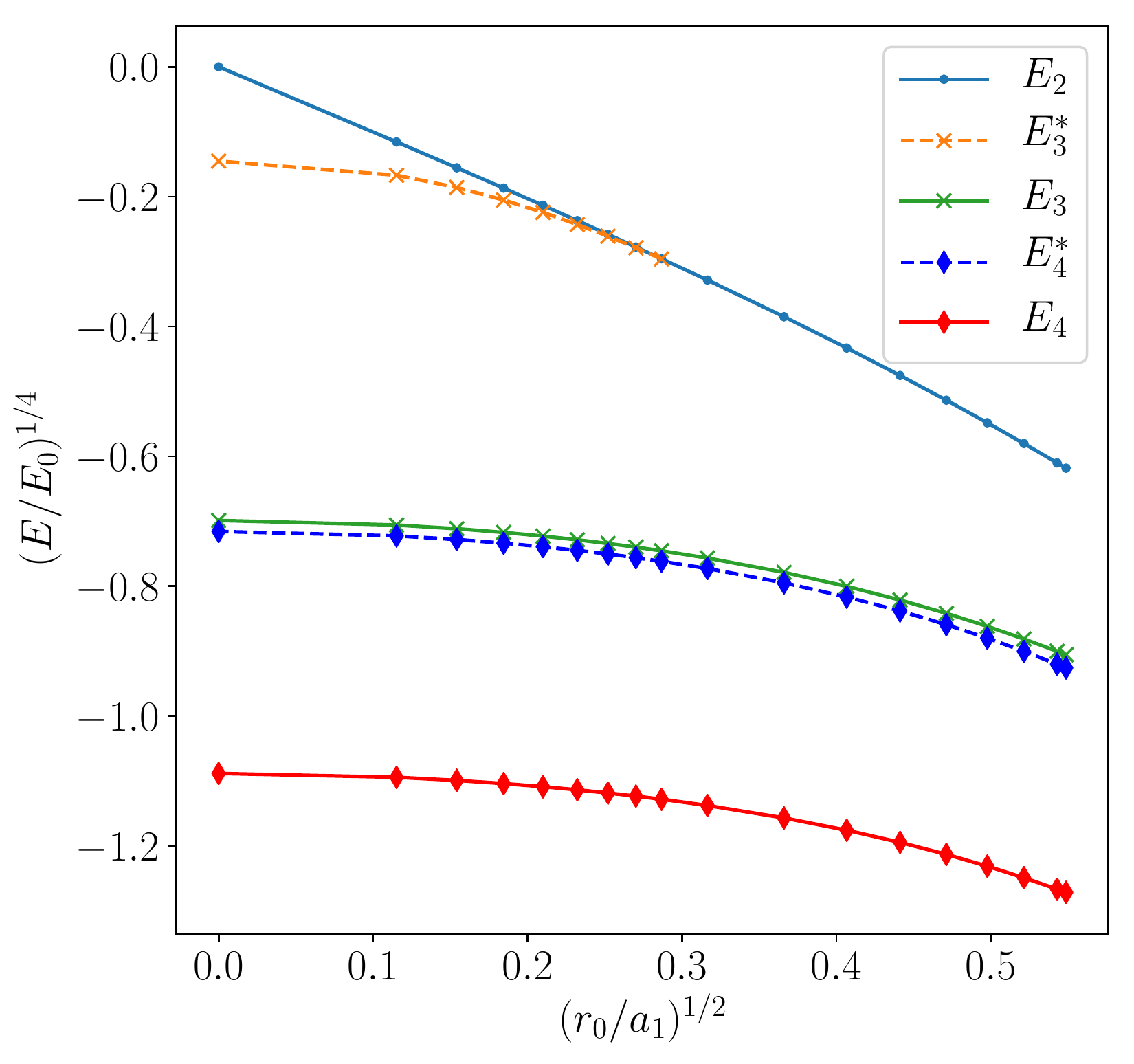}
  \end{center}
  \caption{Efimov plot for $N=2,3,4$ particles along the nuclear
  cut $a_0/a_1=-4.3066$. The triplet scattering length $a_1$ 
  is in units
  of $r_0=1.65$~fm and the energies are expressed in units of 
  $E_0=\hbar^2/mr_0^2 = -15.232$~MeV.}
  \label{fig:efimovPlot}
\end{figure}

From the results reported in Table~\ref{tab:nuclearPlane} we also observe that
the three-body binding energy at the physical point is much larger than the
experimental value of -8.48 MeV; this is a well known fact related to the
necessity of including a three-body force, a point we discuss in the next
section.  In Fig.~\ref{fig:efimovPlot} we show the Efimov plot up to four
particles; we clearly see the three-body excited state disappearing in the
continuum.  We also observe the usual feature of two four-body states attached
to the three-body ground state. The four body calculations are done for the same
quantum numbers as $^4$He, that means $S=0$ and $T=0$, thus the two states have
mostly a symmetric spatial wave function. The ratio between the ground state
energy of the four-body state and the ground state of the three-body state
$E_4/E_3$ is not constant along the path, but it varies from $E_4/E_3 = 5.89$ at
the unitary point to $E_4/E_3 = 3.89$ at the physical point, close to the
realistic case of $3.67$. As far as the excited stated of the four-body system
is concerned, the ratio between its energy and that of the three-body state is
more or less constant along the path $E_4^*/E_3 = 1.09 - 1.1$; the finite-range
corrections result in a bigger value of this ratio with respect to the
zero-range limit~\cite{deltuva:2013_Few-BodySyst}.

\subsection{Universal behavior}

To analyse the universal behavior of the few-nucleon systems we
start recalling the Efimov radial law for three equal
bosons~\cite{braaten:2006_PhysicsReports}
\begin{subequations}
  \begin{equation}
    E_3/E_2 = \tan^2\xi 
  \label{eq:zenergy}
  \end{equation}
  \begin{equation}
    \kappa_*a =
    \text{e}^{(n-n_*)\pi/s_0}\frac{\text{e}^{-\Delta(\xi)/2s_0}}{\cos\xi} \,,
  \label{eq:zkstara}
  \end{equation}
  \label{eq:ZeroRange}
\end{subequations}
where, due to its zero-range character, $E_2=-\hbar^2/ma^2$ and
the three-body binding energy of level $n_*$ at unitary is $\hbar^2/m \kappa_*^2$. 
The function $\Delta(\xi)$ is universal in the sense that it is the same
for all the energy levels. It can be calculated solving the STM equations
as explained  for example  in Ref.~\cite{braaten:2006_PhysicsReports}, and 
its expression can be given in a parametric form~\cite{gattobigio:2019_arXiv}. To be
noticed that the spectrum given by the above equation is not bounded from below. 
For a real three-boson system located close to the unitary limit and interacting through
short-range forces with a typical length $\ell$, the discrete spectrum is bounded from below 
with the number of levels roughly approximate by $(s_0/\pi)\ln(|a|/\ell)$.

The extension of Eqs.~(\ref{eq:ZeroRange}) to describe finite-range interactions,
considering more particles and eventually spin-isospin degrees of freedom,
is given in a series of papers,
Refs.\cite{gattobigio:2014_Phys.Rev.A,gattobigio:2014_JournalofPhysics:ConferenceSeries,kievsky:2014_Phys.Rev.A,kievsky:2016_Few-BodySyst,kievsky:2018_J.Phys.Conf.Ser.},
and it reads 
\begin{subequations}
\begin{equation}
    E^m_A/E_2 = \tan^2\xi 
  \label{eq:energy}
\end{equation}
\begin{equation}
    \kappa^m_A a_B +\Gamma^m_A= \frac{\text{e}^{-\Delta(\xi)/2s_0}}{\cos\xi} \,,
  \label{eq:kstara}
\end{equation}
  \label{eq:FiniteRange}
\end{subequations}
where for three particles, $E^m_3$, $m=0,1,\ldots$, is the energy of the different
branches; in Fig.~\ref{fig:efimovPlot} the first two branches ($m=0,1$) are
shown. For four particles $E^m_4$, $m=0,1$, is the energy of the two states
attached to the lowest three-body branch, $E^0_3$. The length $a_B$ is defined
from the two-body energy as $E_2=-\hbar^2/m a_B^2$. Finally, we have introduced the 
shift parameter, $\Gamma^m_A$, which results almost constant along the unitary
window. 
A recent analysis of the shift parameter for three equal bosons is given in
Ref.~\cite{gattobigio:2019_arXiv}, where it is
related  to
the running three-body parameter introduced in
Ref.~\cite{ji:2015_Phys.Rev.A}.
Eq.~(\ref{eq:kstara}) can also be
written as
\begin{equation}
    \kappa^m_A a_B = \frac{\text{e}^{-\Delta_A^m(\xi)/2s_0}}{\cos\xi} \,,
  \label{eq:kstara1}
\end{equation}
where the shift $\Gamma^m_A$ is absorbed in the level function
$\Delta^m_A(\xi)$; in the present work it is calculated from a Gaussian
potential as in the Bosonic case~\cite{alvarez-rodriguez:2016_Phys.Rev.A}.  In
Ref.~\cite{alvarez-rodriguez:2016_Phys.Rev.A} it has been shown that the level
function $\Delta^m_A(\xi)$, which incorporates the finite-range corrections
given by a Gaussian potential, is about the same for different potentials
close to the unitary limit. Accordingly a Gaussian potential can be thought as a universal 
representation of potential models inside the universality window. Moreover,
the level function $\Delta^m_A(\xi)$ is unique for all Gaussian potentials, it does
not depend on the particular range $r_0$ used for the actual calculations
because, as shown in Fig.~\ref{fig:efimovPlot}, this parameter is just used to
have a dimensionless scattering length and energy. 
This is an important point because the limit $r_0/a_1\rightarrow 0$ can be read either as 
$a_1\rightarrow \infty$ or as $r_0\rightarrow 0$. In the limit $r_0/a_1=0$ the
unitary point coincides with the finite-range-regularized scaling-limit point
and the dimensionless values of the binding momenta are the same for all Gaussian 
potentials. They are given in Tab.~\ref{tab:purenumber} for $A=3,4$ and $m=0,1$.

\begin{table}[t]
  \caption{In this table we report the universal-Gaussian values of the momentum energies of $A=3,4$ 
  systems at the unitary point for the branches $m=(0,1)$, and we summarise the
values of the physical angles, of the Gaussian two-body binding energies
corresponding to the same angle reproduced by the Gaussian potential, and of the
momentum and energy at the unitary limit for the real-nuclear systems
that we have predicted using Eq.~(\ref{eq:equalxi}).}
  \label{tab:purenumber}
\begin{tabular} {@{}c c c c c c c @{}}
\hline\hline
 \rule[-1.2ex]{0pt}{0pt}
 $A$ & $m$ &  $r_0\kappa^m_A\bigr|_G$ & $\tan^2\xi\bigr|_{\text{exp}}$
 & $a_B/r_0\bigr|_G$ & $\kappa^m_A\bigr|_{\text{exp}}$(fm$^{-1}$) & $E^m_A\bigr|_{\text{exp}}$(MeV)\\
 \hline
  3 & 0 & 0.4883 & 3.81 &  2.1866 & 0.2473 & 2.536\\
  3 & 1 & 0.0211 \\
  4 & 0 & 1.1847 & 13.13 & 2.0774 & 0.570 & 13.474\\
  4 & 1 & 0.5124 \\
 \hline
\hline
\end{tabular}
\end{table}

The uniqueness of the Gaussian-level functions and the fact that the Gaussian
potential is an universal representation of potential models close to the
unitary limit, allows us to use the Gaussian potential to predict the values of
the energies at the unitary limit for real systems, which in principle are
described by more realistic potentials. We proceed in the following way: from
Eq.~(\ref{eq:ZeroRange}) we observe that the product $\kappa_* a$ is a function
of the only angle $\xi$ through the universal function $\Delta(\xi)$.  This
property is related to the DSI and it is well verified for real systems, which,
close to the unitary limit, are well represented by the Gaussian level functions
as given in Eq.~(\ref{eq:kstara1}). Therefore, the product $\kappa^m_A a_B$ is
function of solely the angle $\xi$ verifying the following
equality 
\begin{equation}
  \kappa^m_A a_B\Bigr|_{\text{exp}} = \kappa^m_A a_B\Bigr|_{G} \,,
  \label{eq:equalk}
\end{equation}
where $\kappa^m_A a_B\bigr|_{\text{exp}}$ is the function evaluated at the angle given by the
experimental values, and the function $\kappa^m_A a_B\bigr|_G$ is evaluated at the same
angle but calculated with the gaussian potential. From Eq.~(\ref{eq:equalk})
the energy momentum at unitary for the real systems is 
\begin{equation}
  \kappa^m_A\Bigr|_{\text{exp}} = \frac{1}{a_B}\Bigr|_{\text{exp}}\kappa^m_A a_B\Bigr|_G =
  \frac{1}{a_B}\Bigr|_{\text{exp}}(r_0 \kappa^m_A) \frac{a_B}{r_0}\Bigr|_G\,,
  \label{eq:equalxi}
\end{equation}
where the universal Gaussian values of $r_0 \kappa^m_A\bigr|_G$ are reported in 
Table~\ref{tab:purenumber}. 

We can apply Eq.~(\ref{eq:equalxi}) to 
predict the value of the three- and four-body energies at 
the unitary limit for
nuclear physics.
For the three-body case, the experimental binding energies of the deuteron,
$a_B\bigr|_{\text{exp}}=4.3176$~fm, and of the
$^3$H fix the experimental value of the angle $\xi$ to be  $\tan^2\xi\bigr|_{exp} = 3.81$.
Using the range value $r_0=1.65\;$fm, this angle is reproduced by 
the Gaussian strengths $V_{10}=-64.96$~MeV and $V_{01}=-37.4855$~MeV, 
which corresponds to a deuteron energy of $E_2=-3.1858639$~MeV,
or, equivalently, $a_B/r_0\bigr|_G = 2.1866$.
Using the Gaussian value of $r_0\kappa_3^0\bigr|_G = 0.4883$, from
Eq.~(\ref{eq:equalxi}) we obtain
$\kappa^0_3\bigr|_{\text{exp}}=0.2473$~fm$^{-1}$ corresponding to a three-nucleon binding
energy at unitary of $E_3^0\bigr|_\text{exp} = 2.536$~MeV. 

We proceed in the same way for the four-body case. We take $E_4^0=29.1$~MeV as
the experimental value of $^4$He without Coulomb
interaction~\cite{pudliner:1995_Phys.Rev.Lett.}; with this value and that of the
deuteron we obtain $\tan^2\xi\bigr|_{\text{exp}} = 13.1$, which can be reproduced
using the Gaussian potential with $V_{10} = -66.4$~MeV and
$V_{01}=-37.36047$~MeV that also
gives $a_B/r_0\bigr|_G = 2.0774$. Using Eq.~(\ref{eq:equalxi}) and the
universal-Gaussian value $r_0\kappa_4^0\bigr|_G=1.1847$ we obtain 
$\kappa^0_4\bigr|_{\text{exp}}=0.570$~fm$^{-1}$, or, equivalently,
$E_4^0\bigr|_\text{exp} = 13.474$~MeV. All the results are summarised 
in Table~\ref{tab:purenumber}, and it should be noted that 
predictions of the same order exist for $A=3$~\cite{epelbaum:2006_Eur.Phys.J.C}.

\begin{figure}[!tbp]
  \begin{center} 
    \includegraphics[width=\linewidth]{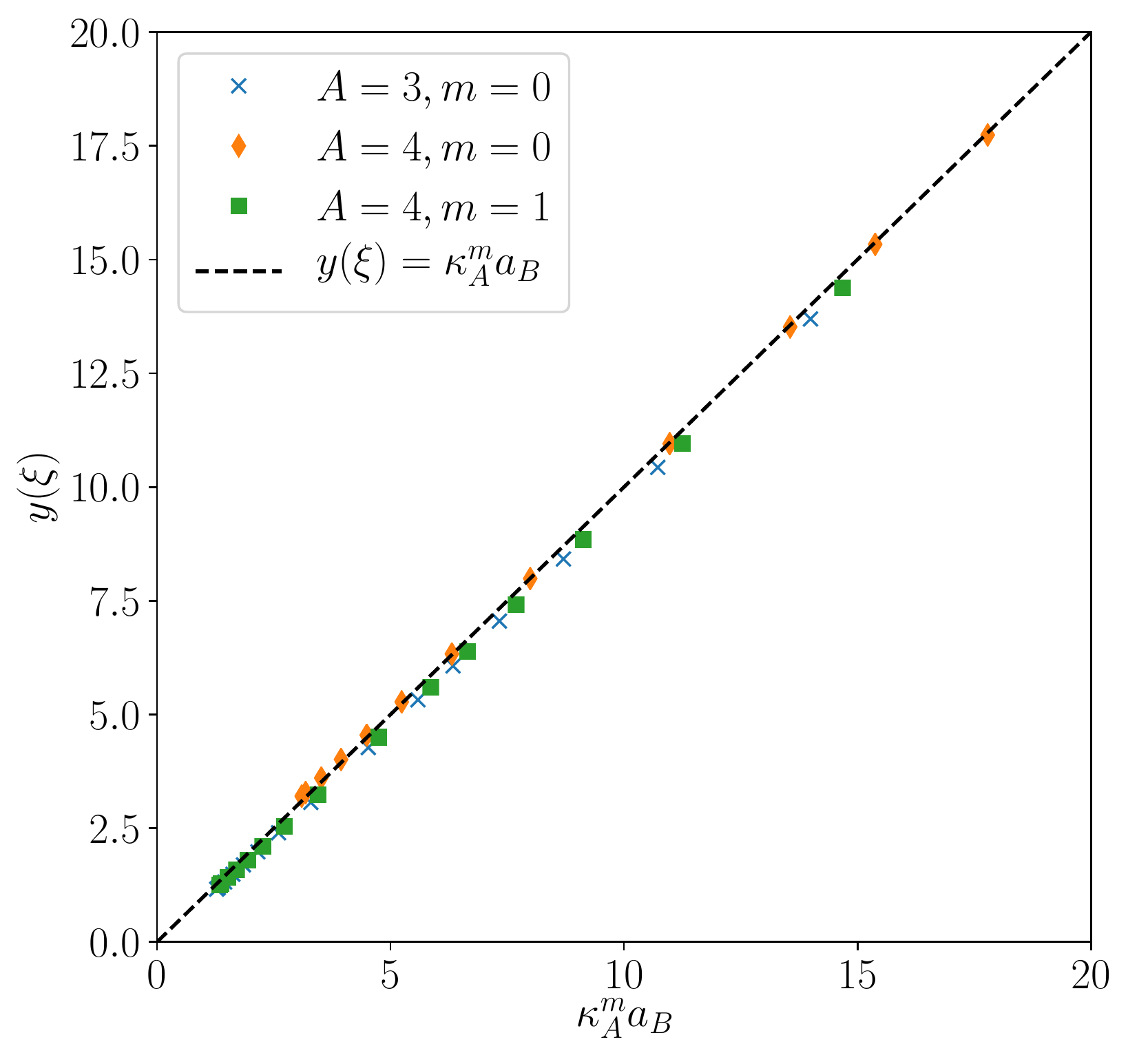}
  \end{center}
  \caption{Efimov plot for the nuclear cut in the form
    of $y(\xi)$, Eq.~(\ref{eq:yofxi}), as a function of  $\kappa_A^m a_B$. The zero-range
  limit is given by the straigth line $y(\xi) = \kappa_A^m a_B$.}
  \label{fig:ylines}
\end{figure}

In order to study further the close relation between the zero-
and finite-range descriptions we look at the behavior of
\begin{equation}
y(\xi) = \text{e}^{-\Delta(\xi)/2s_0}/{\cos\xi} 
  \label{eq:yofxi}
\end{equation}
as a function of $\kappa^m_A a_B$. For  zero-range this function is a
line going through the origin at $45$ degrees. As already
observed~\cite{gattobigio:2014_Phys.Rev.A,alvarez-rodriguez:2016_Phys.Rev.A}
for bosons, if the shift parameter $\Gamma^m_A$ is almost constant,
three and four particles results should give a linear relation between $y(\xi)$ and
$\kappa^m_A a_B$ though not going through the origin. The results are given
in Fig.~\ref{fig:ylines} showing the expected behavior in a very extended range
of $\kappa^m_A a_B$ values.

\subsection{Including the $A=6$ energies}
\begin{figure}
  \begin{center} 
    \includegraphics[width=\linewidth]{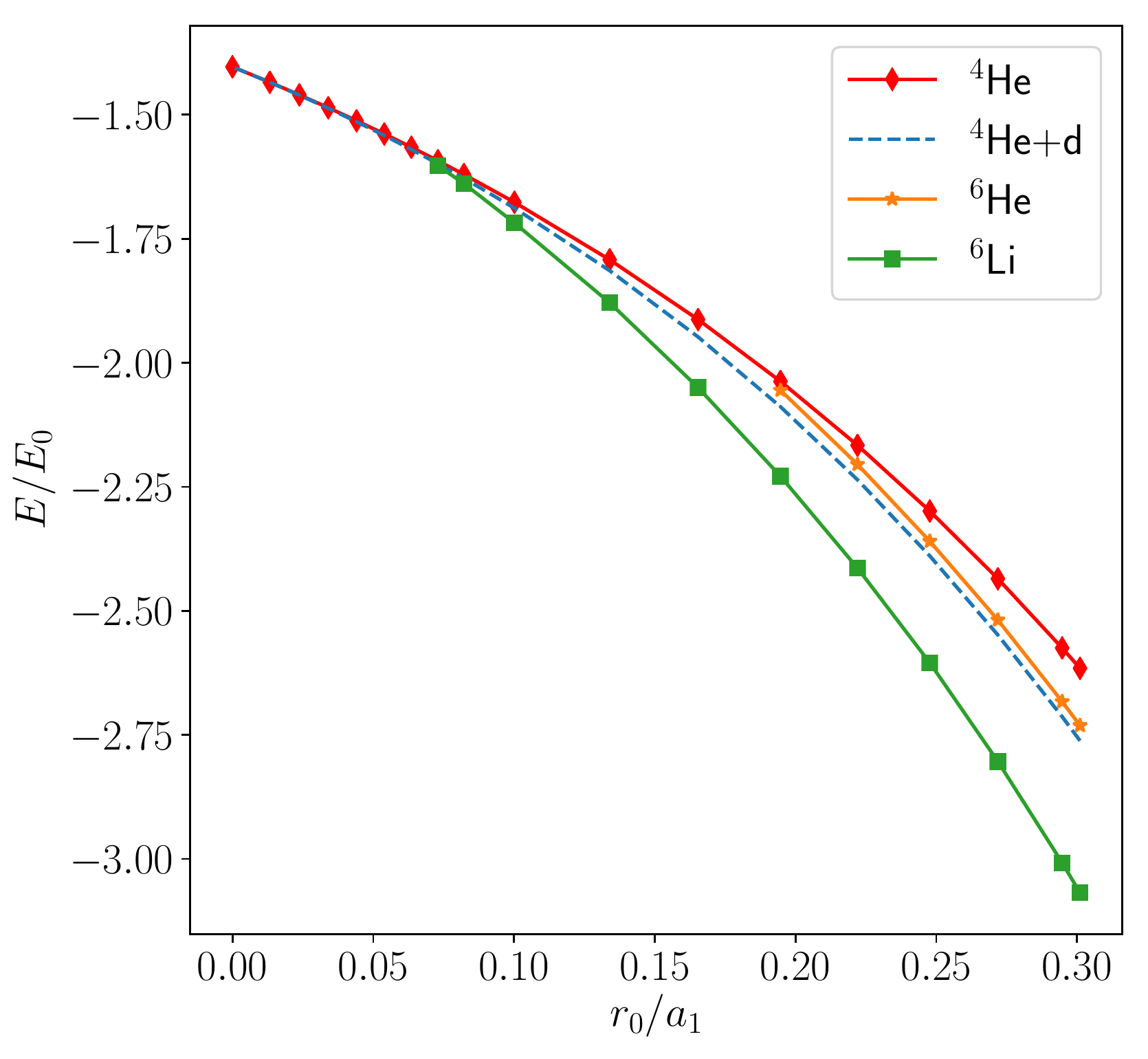}
  \end{center}
  \caption{Efimov plot in the nuclear cut for $A=6$ particles.
   The scattering length is in units of $r_0=1.65$~fm and the 
   energies in units of $E_0=\hbar^2/mr_0^2=15.232$~MeV. We
   distinguish between the six-body state that has the quantum 
   numbers of $^6$He and the one with the quantum numbers of $^6$Li.
   We also report the energy of the $A=4$, which has the quantum number of
   $^4$He, and represents the 
   threshold for the $^6$He, and the energy of $^4$He+d which represents 
   the threshold for $^6$Li. In the present calculations the Coulomb interaction
   has not been taken into account.}
  \label{fig:sixBodyEfimov}
\end{figure}

In the following we study the six-body bound states as a function of the
triplet scattering length along the nuclear cut; we expect a bigger deviation
from the bosonic case because the totally symmetric spatial component 
cannot be anymore present; with only four internal degrees of freedom, the spin
and the isospin, there are only mixed components.
In the $A=6$ case
we distinguish two different states, one with quantum numbers
$S=0$ and $T=1$, to which we refer to as $^6$He even in absence of Coulomb
interaction, and one with quantum numbers $S=1$ and $T=0$, to which
we refer to as $^6$Li. The results of Table~\ref{tab:nuclearPlane} are
reported in Fig.~\ref{fig:sixBodyEfimov}; clearly, we can observe
the absence of these states close to the  unitary limit. 
This is a big difference with respect to the bosonic case,
where, for $6\ge A>3$ the $A$-boson system at unitary has two states, one deep and one shallow,
attached to the $A-1$ ground state
~\cite{gattobigio:2012_Phys.Rev.A,gattobigio:2014_Phys.Rev.A,kievsky:2014_Phys.Rev.A}.
Instead, the two fermionic $A=6$ states are not bound below the $^4$He threshold
(at unitary the $^4$He and $^4$He+d threshold coincide since the two-body system
has zero energy). This is clearly a sign of the absence of the symmetric component in the 
spatial wave function. From the previous discussion we notice the
interesting result that, at the unitary limit, there is a mass gap for $A=5,6$.
This gap continues to exist only for the case $A=5$ at the physical point.

In fact, following the behavior of the $A=6$ states from
Fig.~\ref{fig:sixBodyEfimov} we
observe that, as the two-body system acquires energy, there is a point around
$r_0/a_1\approx0.07$ in which $^6$Li emerges from the $^4$He+d threshold
and, at $r_0/a_1\approx0.2$, $^6$He emerges from the $^4$He threshold.
The difference in energy between the two states at this last point
is of $2.64$~MeV, of the order of the experimental mass difference; it becomes of the order
of $5.14$~MeV at the physical point. We can conclude that this is a  subtle effect of
the finite-range character of the force, as we are going to discuss in the
next sections.

\begin{figure}
  \begin{center} 
    \includegraphics[width=\linewidth]{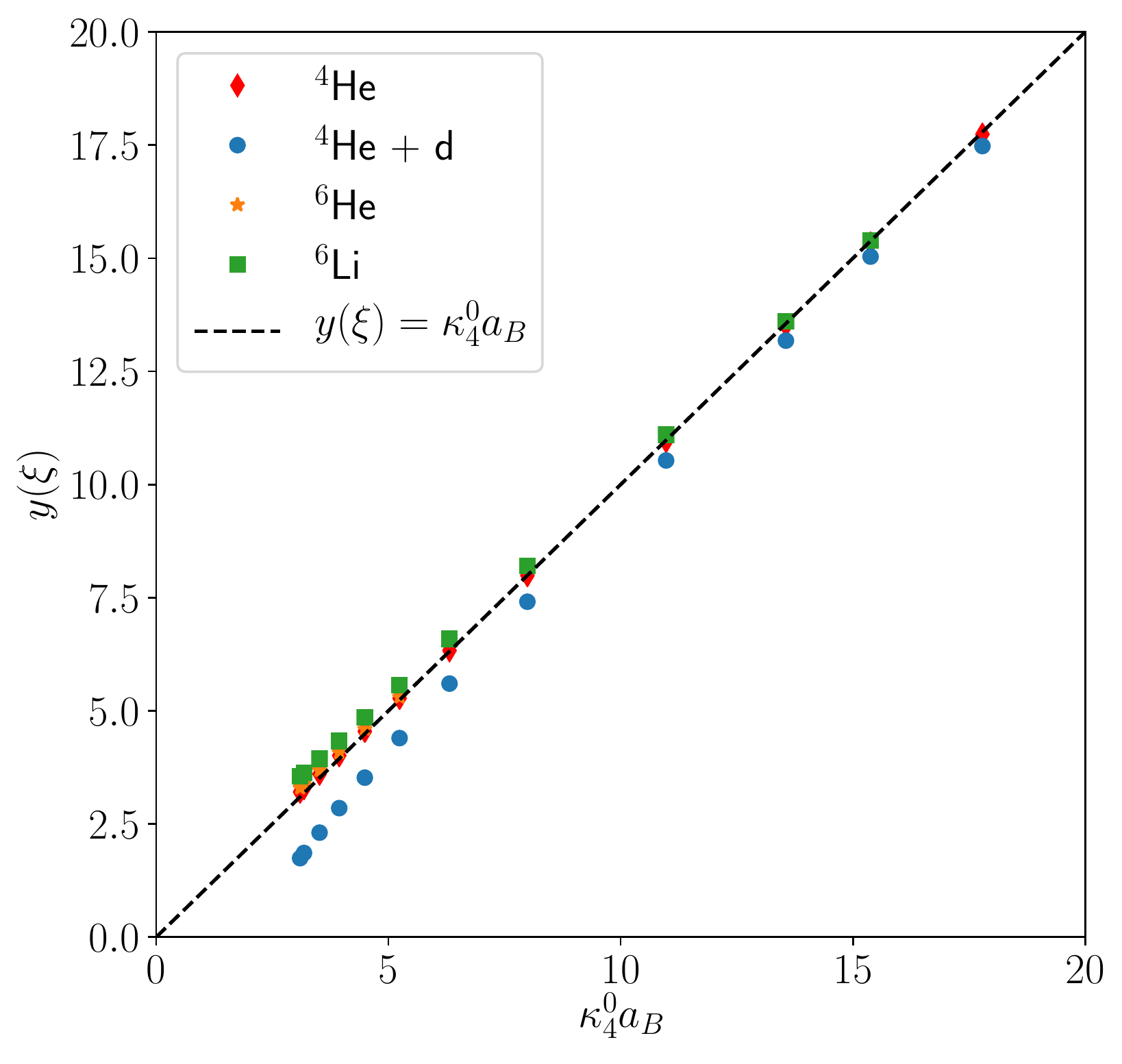}
  \end{center}
  \caption{Efimov plot in the nuclear cut for $A=6$ particles,
  the same as in Fig.~\ref{fig:sixBodyEfimov}, in the form of $y(\xi)$ as a
  function of $\kappa_4^0a_B$. With respect to the $A=3,4$ cases, we observe a
  bigger deviation from the universal prediction of Efimov physics.}
  \label{fig:ylines6}
\end{figure}

Finally, we investigate the universal character of the fermionic $A=6$ states
using Eq.~(\ref{eq:yofxi}). A linear behavior of the function $y(\xi)$
indicates a behavior controlled by the scattering lengths and the three-body
parameter. In Fig.~\ref{fig:ylines6} we plot the value of 
$y(\xi)$, calculated using the $A=6$ energies as a function of
$\kappa_4^0a_B$; the latter has been chosen because, at the unitary point, 
is the energy representing the threshold. 
We find a dominant linear relation close to
the thresholds where the structure of the state is dominated by 
the $^4$He. For $^6$Li deviations from the universal behaviour
appears close to the physical point whereas the $^6$He energies
follow nicely the linear behavior showing a strongly universal character.


\section{The physical point}\label{sec:physicalPoint}
From the calculations of Table~\ref{tab:nuclearPlane} we clearly see that 
the two-body potential Eq.~(\ref{eq:twoBody}) is too simple to
describe the spectrum of light nuclei. On the other hand it captures some
important aspects as the one-level three-nucleon spectrum, the $E_4/E_3$ ratio
and the $A=5$ mass gap. As discussed
in the introduction, the two-body Gaussian potential has to be supplemented
with a three-body potential devised to reproduce the $^3$H energy.
This corresponds, in Efimov physics, to fix the three-body parameter or,
following EFT concepts, the promotion to the LO of the three-body interaction
in order to take into account the unnatural large values
of the scattering lengths.
Here we use an hyper-central three-body potential of the following form
\begin{equation}
  W(\rho)  = W_0\,e^{- (r_{12}^2+r_{13}^2+r_{23}^2)/R_3^2} \,,
  \label{eq:threeBody}
\end{equation}
where $r_{ij}$ is the relative distance between particle $i$ and $j$.  In this
potential there are two-independent parameters, the strength of the
potential $W_0$ and its range $R_3$. In order to reproduce the $^{3}$H binding
energy, $E_{^{3}\text{H}} = -8.482$~MeV, an
infinite number of pairs $(W_0,R_3)$ can be chosen. However a very small number of such pairs (in
fact only two~\cite{kievsky:2016_Few-BodySyst}) reproduce other
physical inputs like the energy of the four-body system or the neutron-deuteron
scattering length $a_{nd}$. 
\begin{table*}
  \caption{Calculation for $A=3,4$ at the physical point, $V_{10}=-60.575$~MeV,
  $V_{01} = -37.9$~MeV, and $E_2=-2.2255$~MeV, for selected 
  three-body force parameters.  In the left part calculations without the Coulomb 
 interaction are reported for $^3$H, $E_4$, and $E_4^*$. In the right part of the 
  table the Coulomb interaction has been included to calculate $^3$He, 
  $^4$He, and the excited state $^4$He$^*$. The latter disappear as bound state
  when the three-body force and the Coulomb interaction are consider together. 
  The experimental values are reported in the last row.}
  \label{tab:physicalPoint}
\begin{tabular} {@{}c c c c c| c c c@{}}
\hline\hline
$W_{0}$(MeV) & $R_3$(fm)  &  $^3$H(MeV) & $E_4$(MeV) & $E_4^*$(MeV)& $^3$He(MeV) & $^4$He(MeV) & $^4$He$^*$(MeV) \\ 
\hline
0      & -     & -10.2455  & -39.843 & -11.193   & -9.426 & -38.789 & -10.655 \\
11.922 & 2.5   & -8.48     & -28.670 & -8.75     & -7.722 & -27.754 &  -      \\
9.072  & 2.8   & -8.48     & -29.014 & -8.79     & -7.718 & -28.060 &  -      \\
7.8    & 3.0   & -8.48     & -29.223 & -8.80     & -7.715 & -28.258 &  -      \\
7.638  & 3.03  & -8.48     & -29.255 & -8.80     & -7.714 & -28.290 &   -      \\ 
7.612  & 3.035 & -8.48     & -29.260 & -8.80     & -7.714 & -28.295 &   -      \\
7.6044 & 3.035 & -8.482    & -29.269 & -8.80     & -7.716 & -28.305 &   -      \\
\cline{3-7}
\multicolumn{2}{c}{Experimental Values} & -8.482 &  & &-7.718  & -28.296& \\
 \hline
\hline
\end{tabular}
\end{table*}

In Table~\ref{tab:physicalPoint} we show selected parameters of the three-body
used to reproduce the energy of $^3$H.  In the left part of the table we report
calculations without Coulomb interaction; we observe the repulsive nature of the
three-body force. Without Coulomb interaction $^3$He is degenerate with $^3$H
and the four body state has an energy $E_4$ lower than the one of $^4$He.
Moreover, the four-body system shows an unphysical excited state $E_4^*$;  this
is the universal Efimov aspect of nuclear physics: for each three-body state
there are two attached four-body states. 
In the right part of Table~\ref{tab:physicalPoint} we show calculations where
the Coulomb interaction has been taken into account. We observe that there are
values of  $R_3$ that allow to reproduce $^3$He, and for these values, the
description of $^4$He is close to the experimental values.

\begin{table}[!tbp]
  \caption{Calculation for $A=3,4$ for the case $W_0=7.6044$~MeV and 
    $R_3 = 3.035$~fm with a slow switch on of Coulomb
    interaction controlled by the parameter $\epsilon$. The threshold of $^3$H+p is $E_{^3\text{H}}=-8.482$~MeV, which 
    implies that the four-body exited state $^4$He$^*$ is no more bounded 
    for $\epsilon\approx 0.75$.}
  \label{tab:coulomb}
\begin{tabular} {@{}c c c c c c c c c@{}}
\hline\hline
$\epsilon$  &  $^3$He(MeV) & $^4$He(MeV) & $^4$He$^*$(MeV) \\ 
\hline
0     &  -8.482  &  -29.269  &  -8.804  \\
0.2   &  -8.327  &  -29.076  &  -8.706  \\
0.4   &  -8.173  &  -28.882  &  -8.618  \\
0.6   &  -8.020  &  -28.689  &  -8.536  \\
0.65  &  -7.982  &  -28.641  &  -8.520  \\
0.7   &  -7.944  &  -28.593  &  -8.501  \\
0.75  &  -7.906  &  -28.545  &  -  \\
0.8   &  -7.868  &  -28.497  &  -  \\
1     &  -7.716  &  -28.305  &  -       \\
\hline
\hline
\end{tabular}
\end{table}
\begin{figure}
  \begin{center} 
    \includegraphics[width=\linewidth]{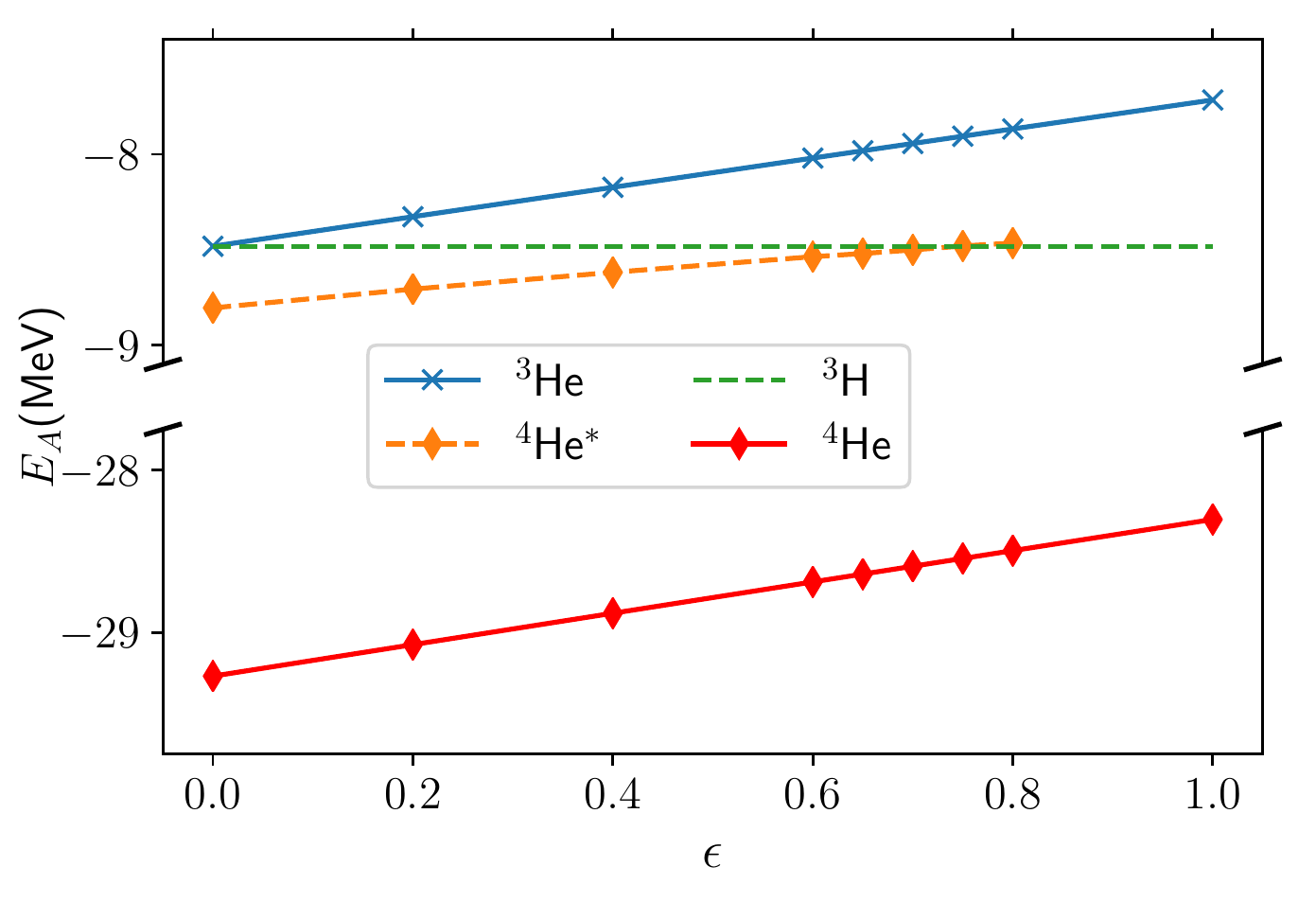}
  \end{center}
  \caption{Evolution of the energies for $A=3,4$ as a function of a smooth
    switching-on of the Coulomb interaction via a multiplicative parameter
    $\epsilon$. The three-body parameters have been fixed to $W_0=7.6044$~MeV
    and $R_3=3.035$~fm. The full Coulomb interaction corresponds to
    $\epsilon=1$.  The four-body excited state disappears for a critical value
    $\epsilon^*=0.754$, while the energies of $^3$He and $^4$He goes to the 
    experimental values better than 0.1\%.}
  \label{fig:coulomb}
\end{figure}

\begin{figure}[!tbp]
  \begin{center} 
    \includegraphics[width=\linewidth]{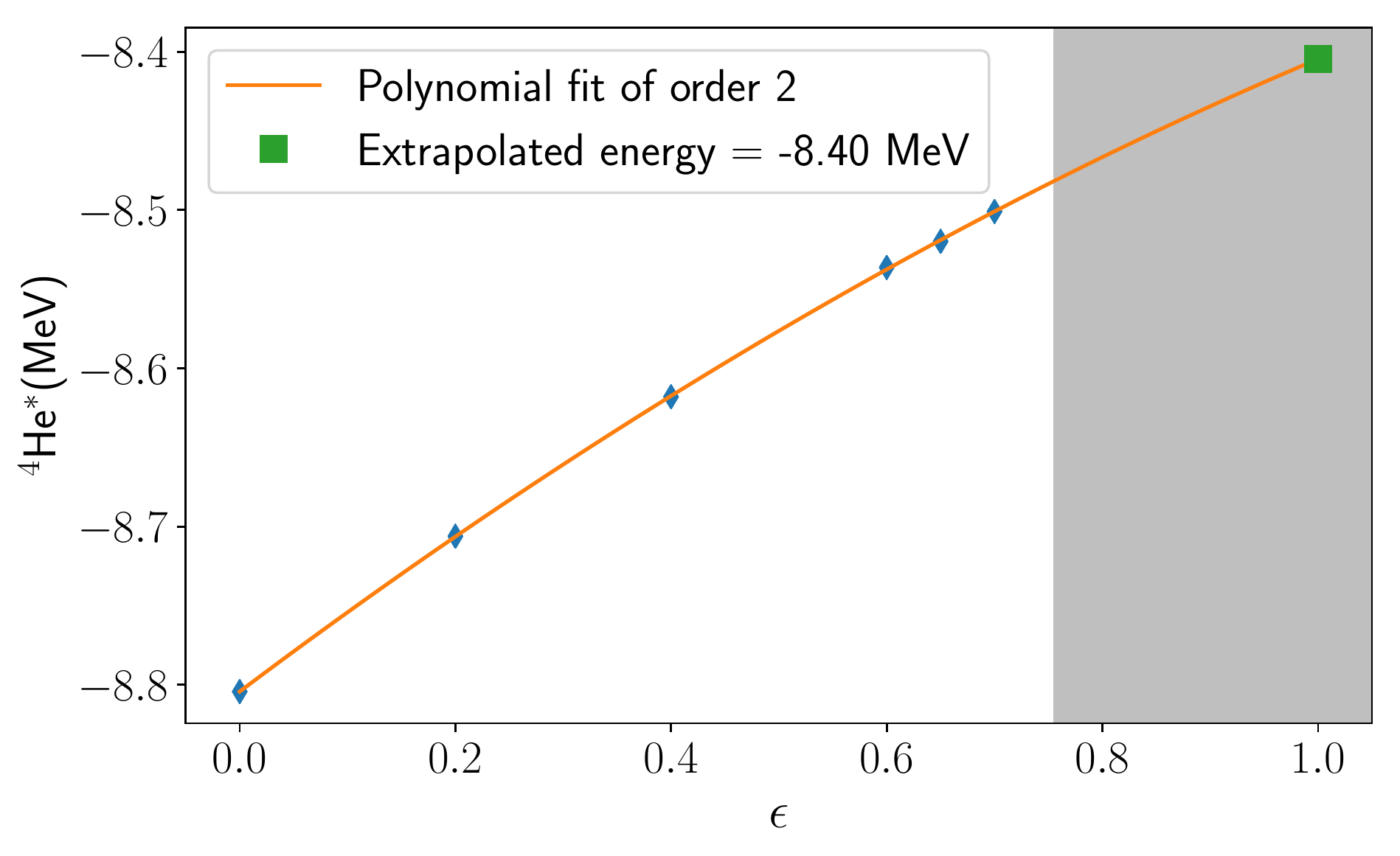}
  \end{center}
  \caption{Energy of the excited four body state $^4$He$^*$ as a function
  of the switching-on of the Coulomb interaction. The grey zone indicates 
  the continuum, which the state enters at $\epsilon^*=0.754$. We extrapolate 
  the state up to full Coulomb $\epsilon=1$, but this does not mean that
  the extrapolated energy corresponds to a resonance because we are not
  taking into account the 
  mixing with the continuum. The experimental position of $0^+$ resonance of $^4$He is 
  -8.086~MeV.}
  \label{fig:extrapolated}
\end{figure}

The presence of Coulomb interaction makes the four-body excited
state disappear. From the table we select the best value of the three-body force,
$W_0=7.6044$~MeV and $R_3=3.035$~fm, to follow the 
evolution of the $^3$He and $^4$He binding energies as a function of a smooth
switching-on of the Coulomb interaction by means of a parameter 
$\epsilon$.
In Table~\ref{tab:coulomb} we report our calculation
as a function of $\epsilon$ and the same data are graphically
represented in Fig.~\ref{fig:coulomb}.
For $\epsilon=0$, that means no Coulomb interaction,
there is only one three-body bound state and the universal two-attached
four-body states. As the value of the Coulomb interaction grows to its
full value, $\epsilon=1$,  the degeneracy between the $^3$H and $^3$He is removed 
and also the value of the ground- and exited-state energy of $^4$He
start to change; for $\epsilon\approx 0.75$ the $^4\text{He}$ excited
state goes behind the $^3$H+p threshold;
a polynomial fit gives the value of threshold at $\epsilon^*=0.754$.
One can probably expect that the fate of this excited state is to become the
known $0^+$ resonance of $^4$He; in order to see this, one should follow the
state as it enters the continuum and mixes with it.  Some preliminary studies do
not support this picture, but there are some indication that the state becomes a
virtual state.  Just as an exercise, we can make a simple extrapolation;
the result of such an exercise is reported in Fig.~\ref{fig:extrapolated}, where
the extrapolated energy is at $-8.40$~MeV, quite far from the experimental
energy of the resonance (-8.0860~MeV). 

To summarise this section, a simple Gaussian-potential acting mainly on $L=0$ 
supplemented with a three-body force and the Coulomb interaction describes
quite accurately the spectrum of light nuclei up to four nucleons. 
The emerging DSI, controlled by the values of the scattering lengths
and the three-nucleon binding energy, strongly constrains the spectrum inside
the universal window. Here we would like to see to which extent the 
energies of $^6$He and $^6$Li are correlated by those parameters.
Though the thresholds are well determined, our observation is that the $L=0$ force, 
even without considering the Coulomb interaction, is not able to bound the 
six-fermion system. 

\section{R\^ole of $P$-waves}\label{sec:pwaves}
\begin{figure}[h]
  \begin{center} 
     \includegraphics[width=8cm]{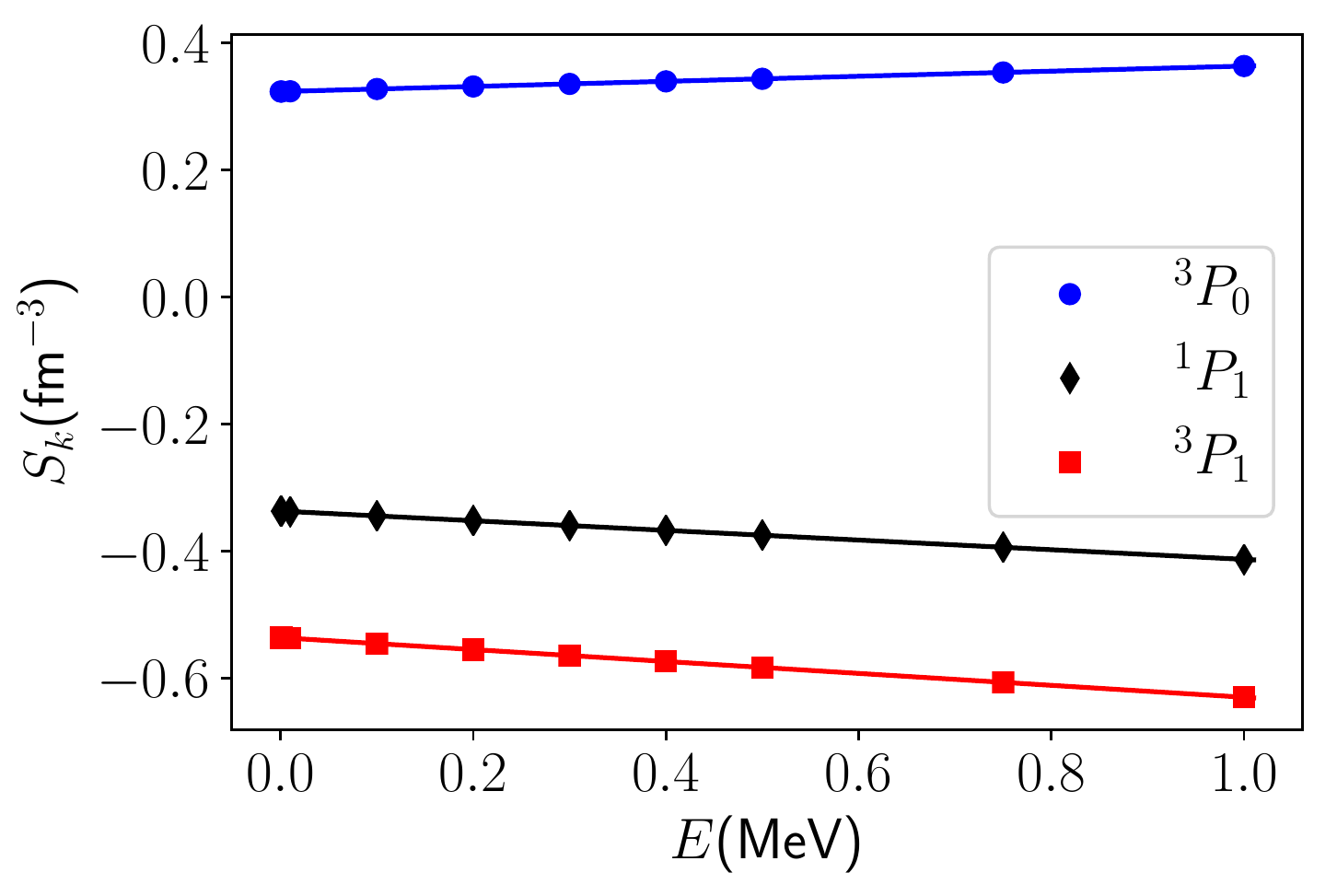}
  \end{center}
  \caption{$P$-wave phase shifts calculated using the AV14 nucleon-nucleon 
  interaction. The points are the effective calculations, while the solid
  lines are fits to that data allowing to extract the scattering parameters, see
  Table~\ref{tab:pskk}.}
  \label{fig:pskk}
\end{figure}

From the previous discussion we have seen that the simple version of the nuclear
interaction dictated by the Efimov physics is not enough to describe the
six-body sector of the light nuclei spectrum. In this section we investigate the
possible r\^ole of the two terms of the potential Eq.~(\ref{eq:twoBody}),
$V_{00}$, and $V_{11}$, that in the previous sections have been set to zero.
These terms contribute to the description of the $P$-waves through the
antisymmetric condition $(-1)^{(L+S+T)}=-1$. At the two-body level the low
energy $P$-wave phase-shifts can be described by an effective range expansion
which, for single channels, is of the form
\begin{equation}
S_k=k^3\cot{}^{2S+1}P_J = \frac{-1}{{}^{2S+1}a_J}+ \frac{1}{2}\; {}^{2S+1}r_J
\,k^2\,,
\label{eq:skk}
\end{equation}
where $^{2S+1}P_J$ is the $P$-wave phase-shift in spin channel $S$ coupled to
total angular momentum $J$, ${}^{2S+1}a_J$ is the scattering volume and
${}^{2S+1}r_J$ is the $P$-wave effective range. In Fig.\ref{fig:pskk} the
effective range function $S_k$ is shown for the uncoupled phases calculated
using the AV14 nucleon-nucleon interaction~\cite{wiringa:1984_Phys.Rev.C}
(circles) together with a fit for those results (solid lines). The linear
behavior is well verified in this energy region and allows to extract the
scattering parameters as given in Table \ref{tab:pskk}.

\begin{table}[h]
  \caption{Scattering parameters of the effective range expansion
    Eq~(\ref{eq:skk}) for the $P$-wave phase shift.}
  \label{tab:pskk}
\begin{tabular} {@{}l c | l c @{}}
\hline\hline
 ${}^{2S+1}a_J$ & [fm$^{-3}$] & ${}^{2S+1}r_J$ & [fm$^{-1}$] \\
\hline
  ${}^{1}a_1$ & 1.437  & ${}^{1}r_1$ & -6.308  \\
  ${}^{3}a_1$ & 1.231  & ${}^{3}r_1$ & -7.786  \\ 
  ${}^{3}a_0$ & -1.457 & ${}^{3}r_0$ &  3.328  \\
 \hline
\hline
\end{tabular}
\end{table}

From the above analysis we can observe that the interaction in channel
$S,T=0,0$ is repulsive whereas the interaction in channel $S,T=1,1$ is 
slightly attractive in $J=0$ wave. In the first case, we reproduce the 
scattering data with the interaction
\begin{equation}
  V_{00} = +1.625~\text{MeV}\quad r_{00} = 4.03~\text{fm}\,;
  \label{eq:v00}
\end{equation}
with this choice, even  ${}^1P_1$ phases are well described. The 
${}^3P_0$ phases are well described with the interaction
\begin{equation}
  V_{11} = -3.857~\text{MeV}\quad r_{11} = 3.35~\text{fm}\,.
  \label{eq:v11}
\end{equation}
However the interaction defined in Eq.(\ref{eq:twoBody}) cannot distinguish between the 
different two-body $J$-states. Accordingly, for the $S,T=1,1$ channel we use a
Gaussian interaction with range $r_{11}=3.35$ and we allow variations of the
strength around the value $-3.857\,$ MeV. We make one step further and we
optimize the interactions in $V_{10}$ and $V_{01}$ to describe the $L=0$ singlet
and triplet scattering lengths and corresponding effective ranges.
The choice for the potentials is the following
\begin{equation}
  \begin{gathered}
  V_{01} = -30.545885~\text{MeV}\quad r_{01} = 1.8310~\text{fm} \\
  V_{10} = -66.5824776~\text{MeV}\quad r_{10} = 1.5579~\text{fm}\,.
\end{gathered}
  \label{eq:newSwaves}
\end{equation}
The potential of Eq.(\ref{eq:twoBody}) is now defined in the four $S,T$
components and, as in the previous calculations, we introduce a three-body force to 
fix the value of the $^3$H. We use two different range $R_3$ to 
explore how the six bodies depend on it.

\begin{table*}[!tbp]
  \caption{For each choice of the $V_{11}$ potential the three-body force has
been tuned to reproduce the energy of the $^3$H. The range of the potential has
been fixed to $r_{11}= 3.35$~fm.}
  \label{tab:pwaves}
\begin{tabular} {@{}c c c c c c c @{}}
\hline\hline
$V_{11}$ (MeV) &  $W_0$ (MeV) & $R_3$ (fm) & $^3$He (MeV) & $^4$He (MeV) & $^6$He (MeV) & $^6$Li (MeV)\\ 
\hline
-3.857   &  7.8375           & 1.4  &  -7.746 & -28.32  &   -30.93   &  -34.86 \\
-3.0     &  7.8104           & 1.4  &  -7.746 & -28.34  &   -29.90   &  -33.67 \\
-3.0     &  13.461           & 1.2  &  -7.749 & -28.20  &   -30.43   &  -34.35 \\ 
-2.5     &  7.7940           & 1.4  &  -7.745 & -28.35  &   -29.25   &  -33.07 \\
-2.5     &  13.433           & 1.2  &  -7.749 & -28.21  &   -29.81   &  -33.63 \\   
-2.0     &  13.405           & 1.2  &  -7.749 & -28.22  &   -29.16   &  -32.93 \\   
-1.78    &  13.392           & 1.2  &  -7.749 & -28.23  &   -28.87   &  -32.64 \\    
\cline{4-7}
\multicolumn{3}{c}{Experimental Values}& -7.718 & -28.296&  -29.268 & -31.9938\\
\hline
\hline
\end{tabular}
\end{table*}

\begin{figure}
  \begin{center} 
    \includegraphics[width=\linewidth]{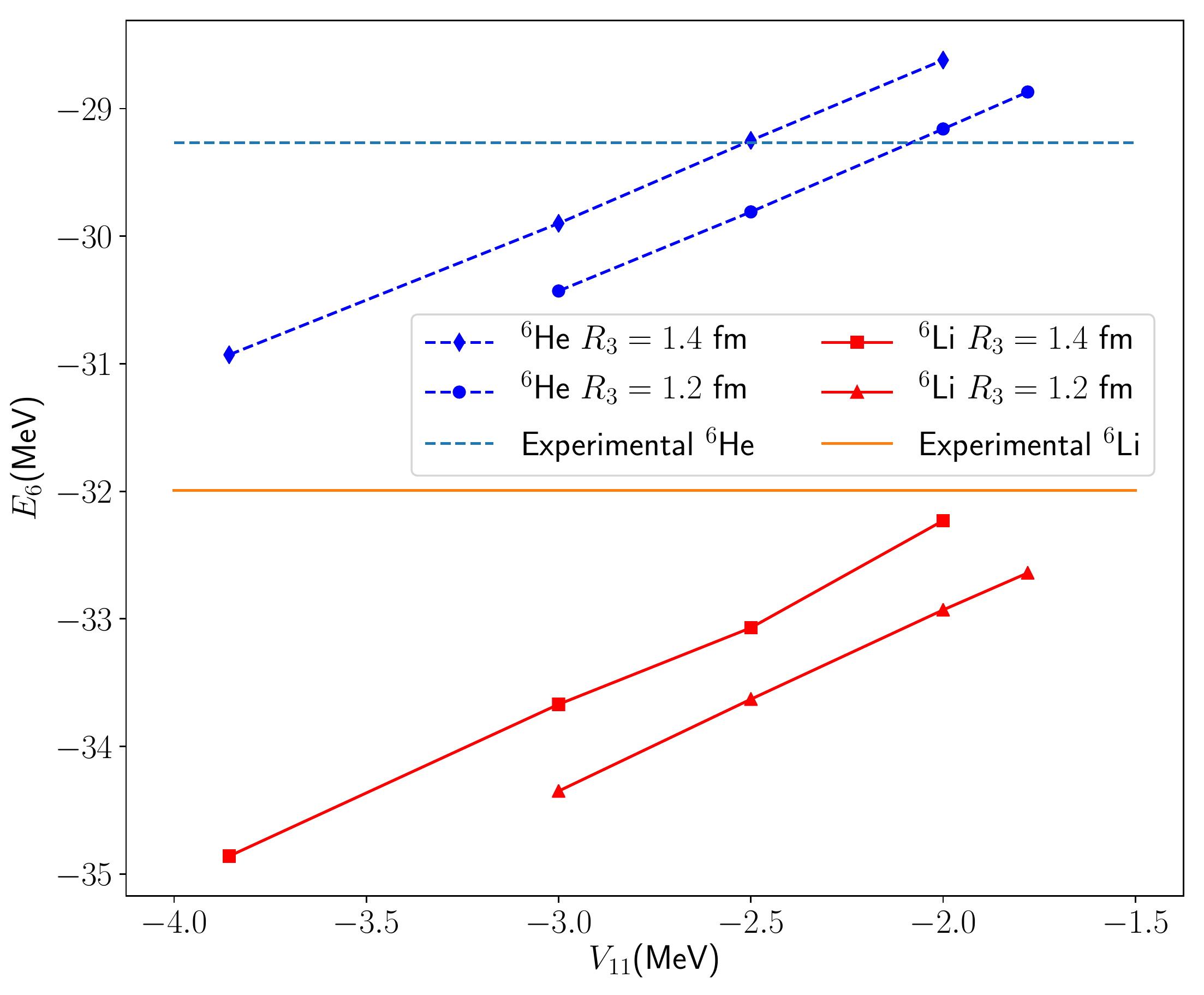}
  \end{center}
  \caption{Energy of $^6$He and $^6$Li as a function of the potential strength
    $V_{11}$. The number on top of each point represent the value of the
  three-body range $R_3$ that has been used. For the sake of comparison, we
also draw the experimental values.}
  \label{fig:pwavesN6}
\end{figure}

In Table~\ref{tab:pwaves} we report our calculations for different choices of
the strength $V_{11}$ and the corresponding three-body strength $W_0$.
In all cases the binding energies of $^3$He and $^4$He are well described
considering that the only charge symmetry breaking component of the force taken
into account is the Coulomb interaction. It is interesting to notice that the
inclusion of the very weak attraction in channel $S,T=1,1$ is enough to bind
 $^6$He and $^6$Li though their bindings are a little bit overpredicted (see
first row of the table). By decreasing the $V_{11}$ strength it is possible to describe
better the $^6$He binding energy, as for example using the strength -2.5 MeV,
but $^6$Li remains overbind by around $1$ MeV. This is a consequence of the lack
of flexibility of the force defined in Eq.(\ref{eq:twoBody}) to distinguish
between the different states in the two-body $P$-channels.
This can be achieved by a spin-orbit term which can remove the 
degeneracy between the three $^3P_J$ phase shifts. In fact the present
interaction predicts a difference $^6$Li - $^6$He mass more or less constant, 1 MeV greater 
than the experimental value.  In Fig.~\ref{fig:pwavesN6} the results for the 
six-body sector are represented; on top of each data we write the range of the
three-body force we used. We observe a linear dependence of the energies with 
respect the strength of $V_{11}$ potential, while the different three-body
ranges just shift the linear dependence. 

\section{Conclusions}\label{sec:conclusions}

The fact that the two $s$-wave scattering lengths, $a_0$ and $a_1$, are large
with respect the natural size of the NN interaction, locates nuclear physics 
inside the universal window. In this context is of interest analyze the
spectrum of $1/2$ spin-isospin fermions controlled by these two parameters.
This very simplified picture has been studied in the first part of the present
work up to six fermion using a Gaussian potential model with variable strength.
Assigning values to the Gaussian strengths in the spin-isospin channels $S,T=0,1$ and $1,0$
the two scattering lengths, $a_0$ and $a_1$, were allowed to vary from infinite
to their physical values following a path, called nuclear cut,
in which the ratio $a_0/a_1=-4.3066$ were kept constant.
Considering only two-body Gaussians and disregarding the Coulomb interaction,
the main results of this analysis are shown in Fig.~\ref{fig:efimovPlot} and Fig.~\ref{fig:sixBodyEfimov}
where the main characteristic of the six fermion spectrum can be seen. At
unitary the $A=5,6$ nuclei are not bound from the $A=4$ threshold. As the system
moved from unitary through the physical point, the tower of infinite three-body
states disappear with only one state surviving. At the same time the six-body
system becomes bound, first the state having the $^6$Li quantum numbers and then
the state having the $^6$He quantum numbers. Moreover all along the path
the excited state of $^4$He is bound with respect to the three-nucleon
threshold. Though the values of the energies are not well reproduced using
a two-body Gaussian interaction, the 
spectrum at the physical point is formed by one two-nucleon state, one three-nucleon state, 
two four-nucleon states and two six-nucleon states.

Two ingredients are missing in this analysis. The first one is trivial and
consists in the inclusion of the Coulomb interaction. The second ingredient
is dictated by EFT concepts and consists in the consideration of a three-body
force. Accordingly, in the second part of the study we concentrate in the
physical point considering those terms in the interaction. The main results
are given in Table~\ref{tab:physicalPoint} where selected parametrization
of the three-body force are shown in order to describe the triton binding
energy. It should be noticed that considering the Coulomb interaction
without including the three-body force or, vice versa, considering the three-body force
without including the Coulomb interaction, produces a four-nucleon spectrum with
two bound states. The three- and four-nucleon spectra go to the correct place
after including both interactions. A detailed study of how the $^4$He$^*$
excited state crosses the threshold to the $^3$H-$p$ continuum is given in 
Fig~\ref{fig:extrapolated}.
Preliminary studies indicate that, with the simply interaction used here, the 
$^4$He$^*$ excited state becomes a virtual state. Furthermore, when both, 
the Coulomb interaction and the three-body force, are taken into account
the two six-fermion states become unbound.

The repulsive character of the three-body force, needed to fix the triton
binding energy, produces a delicate cancellation between the different energy
terms promoting both $^6$Li and $^6$He above the respective thresholds.
In order to see how these two nuclei emerge from their thresholds, in the
final part of this study, we extend the Gaussian potential model to include interactions
in the spin-isospin channels $S,T=0,0$ and $1,1$. The strengths and ranges
of these terms have been fixed to reproduce the NN $P$-wave effective range
expansion, as given for example by the AV14 interaction. Our observation was
that a very weak attractive force in the $S,T=1,1$ channel is sufficient
to bind $^6$Li and $^6$He, however with their mass difference overpredicted
by 1 MeV.

The present analysis supports the picture of an universal window in which the
light nuclear systems are located. To this respect the three control parameters,
the two-scattering lengths and the triton binding energy, fix the spectrum of $A\le4$
nuclei, explain the number of levels, the $A=5$ mass gap and locate the $A=6$ thresholds. 
The very weak binding of the $A=6$ nuclei below the 
$^4$He and $^4$He$+d$ thresholds are due to a weakly attractive $P$-wave interaction. 
A more quantitative description of these weakly bound states
necessitates the consideration of a more complex set of operators in the
interactions as a spin-orbit force. For a similar analysis in the context of
chiral perturbation theory we refer to the recent work~\cite{lu:2018_arXiv}.

%

%

\begin{thebibliography}{41}%
\makeatletter
\providecommand \@ifxundefined [1]{%
 \@ifx{#1\undefined}
}%
\providecommand \@ifnum [1]{%
 \ifnum #1\expandafter \@firstoftwo
 \else \expandafter \@secondoftwo
 \fi
}%
\providecommand \@ifx [1]{%
 \ifx #1\expandafter \@firstoftwo
 \else \expandafter \@secondoftwo
 \fi
}%
\providecommand \natexlab [1]{#1}%
\providecommand \enquote  [1]{``#1''}%
\providecommand \bibnamefont  [1]{#1}%
\providecommand \bibfnamefont [1]{#1}%
\providecommand \citenamefont [1]{#1}%
\providecommand \href@noop [0]{\@secondoftwo}%
\providecommand \href [0]{\begingroup \@sanitize@url \@href}%
\providecommand \@href[1]{\@@startlink{#1}\@@href}%
\providecommand \@@href[1]{\endgroup#1\@@endlink}%
\providecommand \@sanitize@url [0]{\catcode `\\12\catcode `\$12\catcode
  `\&12\catcode `\#12\catcode `\^12\catcode `\_12\catcode `\%12\relax}%
\providecommand \@@startlink[1]{}%
\providecommand \@@endlink[0]{}%
\providecommand \url  [0]{\begingroup\@sanitize@url \@url }%
\providecommand \@url [1]{\endgroup\@href {#1}{\urlprefix }}%
\providecommand \urlprefix  [0]{URL }%
\providecommand \Eprint [0]{\href }%
\providecommand \doibase [0]{http://dx.doi.org/}%
\providecommand \selectlanguage [0]{\@gobble}%
\providecommand \bibinfo  [0]{\@secondoftwo}%
\providecommand \bibfield  [0]{\@secondoftwo}%
\providecommand \translation [1]{[#1]}%
\providecommand \BibitemOpen [0]{}%
\providecommand \bibitemStop [0]{}%
\providecommand \bibitemNoStop [0]{.\EOS\space}%
\providecommand \EOS [0]{\spacefactor3000\relax}%
\providecommand \BibitemShut  [1]{\csname bibitem#1\endcsname}%
\let\auto@bib@innerbib\@empty
\bibitem [{\citenamefont {Braaten}\ and\ \citenamefont
  {Hammer}(2006)}]{braaten:2006_PhysicsReports}%
  \BibitemOpen
  \bibfield  {author} {\bibinfo {author} {\bibfnamefont {Eric}\ \bibnamefont
  {Braaten}}\ and\ \bibinfo {author} {\bibfnamefont {H.-W.}\ \bibnamefont
  {Hammer}},\ }\bibfield  {title} {\enquote {\bibinfo {title} {Universality in
  few-body systems with large scattering length},}\ }\href {\doibase
  10.1016/j.physrep.2006.03.001} {\bibfield  {journal} {\bibinfo  {journal}
  {Physics Reports}\ }\textbf {\bibinfo {volume} {428}},\ \bibinfo {pages}
  {259--390} (\bibinfo {year} {2006})}\BibitemShut {NoStop}%
\bibitem [{\citenamefont {Efimov}(1970)}]{efimov:1970_Phys.Lett.B}%
  \BibitemOpen
  \bibfield  {author} {\bibinfo {author} {\bibfnamefont {V}~\bibnamefont
  {Efimov}},\ }\bibfield  {title} {\enquote {\bibinfo {title} {Energy levels
  arising from resonant two-body forces in a three-body system},}\ }\href
  {\doibase 10.1016/0370-2693(70)90349-7} {\bibfield  {journal} {\bibinfo
  {journal} {Phys. Lett. B}\ }\textbf {\bibinfo {volume} {33}},\ \bibinfo
  {pages} {563--564} (\bibinfo {year} {1970})}\BibitemShut {NoStop}%
\bibitem [{\citenamefont {Efimov}(1971)}]{efimov:1971_Sov.J.Nucl.Phys.}%
  \BibitemOpen
  \bibfield  {author} {\bibinfo {author} {\bibfnamefont {V}~\bibnamefont
  {Efimov}},\ }\bibfield  {title} {\enquote {\bibinfo {title} {Weak {{Bound
  States}} of {{Three Resonantly Interacting Particles}}},}\ }\href@noop {}
  {\bibfield  {journal} {\bibinfo  {journal} {Sov. J. Nucl. Phys.}\ }\textbf
  {\bibinfo {volume} {12}},\ \bibinfo {pages} {589} (\bibinfo {year} {1971})},\
  \bibinfo {note} {[Yad. Fiz. 12, 1080\textendash{}1090 (1970)].}\BibitemShut
  {Stop}%
\bibitem [{\citenamefont {Chin}\ \emph {et~al.}(2010)\citenamefont {Chin},
  \citenamefont {{Grimm, Rudolf}}, \citenamefont {Julienne},\ and\
  \citenamefont {Tiesinga}}]{chin:2010_Rev.Mod.Phys.}%
  \BibitemOpen
  \bibfield  {author} {\bibinfo {author} {\bibfnamefont {Cheng}\ \bibnamefont
  {Chin}}, \bibinfo {author} {\bibnamefont {{Grimm, Rudolf}}}, \bibinfo
  {author} {\bibfnamefont {Paul}\ \bibnamefont {Julienne}}, \ and\ \bibinfo
  {author} {\bibfnamefont {Eite}\ \bibnamefont {Tiesinga}},\ }\bibfield
  {title} {\enquote {\bibinfo {title} {Feshbach resonances in ultracold
  gases},}\ }\href {\doibase 10.1103/RevModPhys.82.1225} {\bibfield  {journal}
  {\bibinfo  {journal} {Rev. Mod. Phys.}\ }\textbf {\bibinfo {volume} {82}},\
  \bibinfo {pages} {1225--1286} (\bibinfo {year} {2010})}\BibitemShut {NoStop}%
\bibitem [{\citenamefont {Luo}\ \emph {et~al.}(1993)\citenamefont {Luo},
  \citenamefont {McBane}, \citenamefont {Kim}, \citenamefont {Giese},\ and\
  \citenamefont {Gentry}}]{luo:1993_J.Chem.Phys.}%
  \BibitemOpen
  \bibfield  {author} {\bibinfo {author} {\bibfnamefont {Fei}\ \bibnamefont
  {Luo}}, \bibinfo {author} {\bibfnamefont {George~C.}\ \bibnamefont {McBane}},
  \bibinfo {author} {\bibfnamefont {Geunsik}\ \bibnamefont {Kim}}, \bibinfo
  {author} {\bibfnamefont {Clayton~F.}\ \bibnamefont {Giese}}, \ and\ \bibinfo
  {author} {\bibfnamefont {W.~Ronald}\ \bibnamefont {Gentry}},\ }\bibfield
  {title} {\enquote {\bibinfo {title} {The weakest bond: {{Experimental}}
  observation of helium dimer},}\ }\href {\doibase 10.1063/1.464079} {\bibfield
   {journal} {\bibinfo  {journal} {J. Chem. Phys.}\ }\textbf {\bibinfo {volume}
  {98}},\ \bibinfo {pages} {3564} (\bibinfo {year} {1993})}\BibitemShut
  {NoStop}%
\bibitem [{\citenamefont {Wigner}(1933)}]{wigner:1933_Phys.Rev.}%
  \BibitemOpen
  \bibfield  {author} {\bibinfo {author} {\bibfnamefont {E.}~\bibnamefont
  {Wigner}},\ }\bibfield  {title} {\enquote {\bibinfo {title} {On the {{Mass
  Defect}} of {{Helium}}},}\ }\href {\doibase 10.1103/PhysRev.43.252}
  {\bibfield  {journal} {\bibinfo  {journal} {Phys. Rev.}\ }\textbf {\bibinfo
  {volume} {43}},\ \bibinfo {pages} {252--257} (\bibinfo {year}
  {1933})}\BibitemShut {NoStop}%
\bibitem [{\citenamefont {Fermi}(1936)}]{fermi:1936_Ricercasci.}%
  \BibitemOpen
  \bibfield  {author} {\bibinfo {author} {\bibfnamefont {Enrico}\ \bibnamefont
  {Fermi}},\ }\bibfield  {title} {\enquote {\bibinfo {title} {Sul moto dei
  neutroni nelle sostanze idrogenate},}\ }\href@noop {} {\bibfield  {journal}
  {\bibinfo  {journal} {Ricerca sci.}\ }\textbf {\bibinfo {volume} {7}},\
  \bibinfo {pages} {13--52} (\bibinfo {year} {1936})}\BibitemShut {NoStop}%
\bibitem [{\citenamefont {Huang}\ and\ \citenamefont
  {Yang}(1957)}]{huang:1957_Phys.Rev.}%
  \BibitemOpen
  \bibfield  {author} {\bibinfo {author} {\bibfnamefont {Kerson}\ \bibnamefont
  {Huang}}\ and\ \bibinfo {author} {\bibfnamefont {C.~N.}\ \bibnamefont
  {Yang}},\ }\bibfield  {title} {\enquote {\bibinfo {title}
  {Quantum-{{Mechanical Many}}-{{Body Problem}} with {{Hard}}-{{Sphere
  Interaction}}},}\ }\href {\doibase 10.1103/PhysRev.105.767} {\bibfield
  {journal} {\bibinfo  {journal} {Phys. Rev.}\ }\textbf {\bibinfo {volume}
  {105}},\ \bibinfo {pages} {767--775} (\bibinfo {year} {1957})}\BibitemShut
  {NoStop}%
\bibitem [{\citenamefont {Braaten}\ and\ \citenamefont
  {Hammer}(2003)}]{braaten:2003_Phys.Rev.Lett.}%
  \BibitemOpen
  \bibfield  {author} {\bibinfo {author} {\bibfnamefont {Eric}\ \bibnamefont
  {Braaten}}\ and\ \bibinfo {author} {\bibfnamefont {H.-W.}\ \bibnamefont
  {Hammer}},\ }\bibfield  {title} {\enquote {\bibinfo {title} {An {{Infrared
  Renormalization Group Limit Cycle}} in {{QCD}}},}\ }\href {\doibase
  10.1103/PhysRevLett.91.102002} {\bibfield  {journal} {\bibinfo  {journal}
  {Phys. Rev. Lett.}\ }\textbf {\bibinfo {volume} {91}},\ \bibinfo {pages}
  {102002} (\bibinfo {year} {2003})}\BibitemShut {NoStop}%
\bibitem [{\citenamefont {Epelbaum}\ \emph {et~al.}(2006)\citenamefont
  {Epelbaum}, \citenamefont {Hammer}, \citenamefont {Mei\ss{}ner},\ and\
  \citenamefont {Nogga}}]{epelbaum:2006_Eur.Phys.J.C}%
  \BibitemOpen
  \bibfield  {author} {\bibinfo {author} {\bibfnamefont {E.}~\bibnamefont
  {Epelbaum}}, \bibinfo {author} {\bibfnamefont {H.-W.}\ \bibnamefont
  {Hammer}}, \bibinfo {author} {\bibfnamefont {U.-G.}\ \bibnamefont
  {Mei\ss{}ner}}, \ and\ \bibinfo {author} {\bibfnamefont {A.}~\bibnamefont
  {Nogga}},\ }\bibfield  {title} {\enquote {\bibinfo {title} {More on the
  infrared renormalization group limit cycle in {{QCD}}},}\ }\href {\doibase
  10.1140/epjc/s10052-006-0004-x} {\bibfield  {journal} {\bibinfo  {journal}
  {Eur. Phys. J. C}\ }\textbf {\bibinfo {volume} {48}},\ \bibinfo {pages}
  {169--178} (\bibinfo {year} {2006})}\BibitemShut {NoStop}%
\bibitem [{\citenamefont {Beane}\ \emph {et~al.}(2002)\citenamefont {Beane},
  \citenamefont {Bedaque}, \citenamefont {Savage},\ and\ \citenamefont {{van
  Kolck}}}]{beane:2002_Nucl.Phys.A}%
  \BibitemOpen
  \bibfield  {author} {\bibinfo {author} {\bibfnamefont {S.R.}\ \bibnamefont
  {Beane}}, \bibinfo {author} {\bibfnamefont {P.F.}\ \bibnamefont {Bedaque}},
  \bibinfo {author} {\bibfnamefont {M.J.}\ \bibnamefont {Savage}}, \ and\
  \bibinfo {author} {\bibfnamefont {U.}~\bibnamefont {{van Kolck}}},\
  }\bibfield  {title} {\enquote {\bibinfo {title} {Towards a perturbative
  theory of nuclear forces},}\ }\href {\doibase 10.1016/S0375-9474(01)01324-0}
  {\bibfield  {journal} {\bibinfo  {journal} {Nucl. Phys. A}\ }\textbf
  {\bibinfo {volume} {700}},\ \bibinfo {pages} {377--402} (\bibinfo {year}
  {2002})}\BibitemShut {NoStop}%
\bibitem [{\citenamefont {{van Kolck}}(1999)}]{vankolck:1999_NuclearPhysicsA}%
  \BibitemOpen
  \bibfield  {author} {\bibinfo {author} {\bibfnamefont {U.}~\bibnamefont {{van
  Kolck}}},\ }\bibfield  {title} {\enquote {\bibinfo {title} {Effective field
  theory of short-range forces},}\ }\href {\doibase
  10.1016/S0375-9474(98)00612-5} {\bibfield  {journal} {\bibinfo  {journal}
  {Nuclear Physics A}\ }\textbf {\bibinfo {volume} {645}},\ \bibinfo {pages}
  {273--302} (\bibinfo {year} {1999})}\BibitemShut {NoStop}%
\bibitem [{\citenamefont {Bedaque}\ \emph
  {et~al.}(1999{\natexlab{a}})\citenamefont {Bedaque}, \citenamefont {Hammer},\
  and\ \citenamefont {{van Kolck}}}]{bedaque:1999_Phys.Rev.Lett.}%
  \BibitemOpen
  \bibfield  {author} {\bibinfo {author} {\bibfnamefont {P.}~\bibnamefont
  {Bedaque}}, \bibinfo {author} {\bibfnamefont {H.-W.}\ \bibnamefont {Hammer}},
  \ and\ \bibinfo {author} {\bibfnamefont {U.}~\bibnamefont {{van Kolck}}},\
  }\bibfield  {title} {\enquote {\bibinfo {title} {Renormalization of the
  {{Three}}-{{Body System}} with {{Short}}-{{Range Interactions}}},}\ }\href
  {\doibase 10.1103/PhysRevLett.82.463} {\bibfield  {journal} {\bibinfo
  {journal} {Phys. Rev. Lett.}\ }\textbf {\bibinfo {volume} {82}},\ \bibinfo
  {pages} {463--467} (\bibinfo {year} {1999}{\natexlab{a}})}\BibitemShut
  {NoStop}%
\bibitem [{\citenamefont {Bedaque}\ \emph
  {et~al.}(1999{\natexlab{b}})\citenamefont {Bedaque}, \citenamefont {Hammer},\
  and\ \citenamefont {van Kolck}}]{bedaque:1999_Nucl.Phys.A}%
  \BibitemOpen
  \bibfield  {author} {\bibinfo {author} {\bibfnamefont {P.~F.}\ \bibnamefont
  {Bedaque}}, \bibinfo {author} {\bibfnamefont {H.~W.}\ \bibnamefont {Hammer}},
  \ and\ \bibinfo {author} {\bibfnamefont {U.}~\bibnamefont {van Kolck}},\
  }\bibfield  {title} {\enquote {\bibinfo {title} {The three-boson system with
  short-range interactions},}\ }\href {\doibase DOI:
  10.1016/S0375-9474(98)00650-2} {\bibfield  {journal} {\bibinfo  {journal}
  {Nucl. Phys. A}\ }\textbf {\bibinfo {volume} {646}},\ \bibinfo {pages} {444
  -- 466} (\bibinfo {year} {1999}{\natexlab{b}})}\BibitemShut {NoStop}%
\bibitem [{\citenamefont
  {Grie\ss{}hammer}(2005)}]{griesshammer:2005_Nucl.Phys.A}%
  \BibitemOpen
  \bibfield  {author} {\bibinfo {author} {\bibfnamefont {Harald~W.}\
  \bibnamefont {Grie\ss{}hammer}},\ }\bibfield  {title} {\enquote {\bibinfo
  {title} {Na\"ive dimensional analysis for three-body forces without pions},}\
  }\href {\doibase 10.1016/j.nuclphysa.2005.05.202} {\bibfield  {journal}
  {\bibinfo  {journal} {Nucl. Phys. A}\ }\textbf {\bibinfo {volume} {760}},\
  \bibinfo {pages} {110--138} (\bibinfo {year} {2005})}\BibitemShut {NoStop}%
\bibitem [{\citenamefont {Kievsky}\ \emph
  {et~al.}(2017{\natexlab{a}})\citenamefont {Kievsky}, \citenamefont {Viviani},
  \citenamefont {Gattobigio},\ and\ \citenamefont
  {Girlanda}}]{kievsky:2017_Phys.Rev.C}%
  \BibitemOpen
  \bibfield  {author} {\bibinfo {author} {\bibfnamefont {A.}~\bibnamefont
  {Kievsky}}, \bibinfo {author} {\bibfnamefont {M.}~\bibnamefont {Viviani}},
  \bibinfo {author} {\bibfnamefont {M.}~\bibnamefont {Gattobigio}}, \ and\
  \bibinfo {author} {\bibfnamefont {L.}~\bibnamefont {Girlanda}},\ }\bibfield
  {title} {\enquote {\bibinfo {title} {Implications of {{Efimov}} physics for
  the description of three and four nucleons in chiral effective field
  theory},}\ }\href {\doibase 10.1103/PhysRevC.95.024001} {\bibfield  {journal}
  {\bibinfo  {journal} {Phys. Rev. C}\ }\textbf {\bibinfo {volume} {95}}
  (\bibinfo {year} {2017}{\natexlab{a}}),\
  10.1103/PhysRevC.95.024001}\BibitemShut {NoStop}%
\bibitem [{\citenamefont {Gattobigio}\ \emph
  {et~al.}(2011{\natexlab{a}})\citenamefont {Gattobigio}, \citenamefont
  {Kievsky},\ and\ \citenamefont {Viviani}}]{gattobigio:2011_Phys.Rev.A}%
  \BibitemOpen
  \bibfield  {author} {\bibinfo {author} {\bibfnamefont {M.}~\bibnamefont
  {Gattobigio}}, \bibinfo {author} {\bibfnamefont {A.}~\bibnamefont {Kievsky}},
  \ and\ \bibinfo {author} {\bibfnamefont {M.}~\bibnamefont {Viviani}},\
  }\bibfield  {title} {\enquote {\bibinfo {title} {Spectra of helium clusters
  with up to six atoms using soft-core potentials},}\ }\href {\doibase
  10.1103/PhysRevA.84.052503} {\bibfield  {journal} {\bibinfo  {journal} {Phys.
  Rev. A}\ }\textbf {\bibinfo {volume} {84}},\ \bibinfo {pages} {052503}
  (\bibinfo {year} {2011}{\natexlab{a}})}\BibitemShut {NoStop}%
\bibitem [{\citenamefont {Kievsky}\ \emph {et~al.}(2014)\citenamefont
  {Kievsky}, \citenamefont {Timofeyuk},\ and\ \citenamefont
  {Gattobigio}}]{kievsky:2014_Phys.Rev.A}%
  \BibitemOpen
  \bibfield  {author} {\bibinfo {author} {\bibfnamefont {A.}~\bibnamefont
  {Kievsky}}, \bibinfo {author} {\bibfnamefont {N.~K.}\ \bibnamefont
  {Timofeyuk}}, \ and\ \bibinfo {author} {\bibfnamefont {M.}~\bibnamefont
  {Gattobigio}},\ }\bibfield  {title} {\enquote {\bibinfo {title}
  {\${{N}}\$-boson spectrum from a discrete scale invariance},}\ }\href
  {\doibase 10.1103/PhysRevA.90.032504} {\bibfield  {journal} {\bibinfo
  {journal} {Phys. Rev. A}\ }\textbf {\bibinfo {volume} {90}},\ \bibinfo
  {pages} {032504} (\bibinfo {year} {2014})}\BibitemShut {NoStop}%
\bibitem [{\citenamefont {Kievsky}\ \emph
  {et~al.}(2017{\natexlab{b}})\citenamefont {Kievsky}, \citenamefont {Polls},
  \citenamefont {{Juli\'a-D\'iaz}},\ and\ \citenamefont
  {Timofeyuk}}]{kievsky:2017_Phys.Rev.A}%
  \BibitemOpen
  \bibfield  {author} {\bibinfo {author} {\bibfnamefont {A.}~\bibnamefont
  {Kievsky}}, \bibinfo {author} {\bibfnamefont {A.}~\bibnamefont {Polls}},
  \bibinfo {author} {\bibfnamefont {B.}~\bibnamefont {{Juli\'a-D\'iaz}}}, \
  and\ \bibinfo {author} {\bibfnamefont {N.~K.}\ \bibnamefont {Timofeyuk}},\
  }\bibfield  {title} {\enquote {\bibinfo {title} {Saturation properties of
  helium drops from a leading-order description},}\ }\href {\doibase
  10.1103/PhysRevA.96.040501} {\bibfield  {journal} {\bibinfo  {journal} {Phys.
  Rev. A}\ }\textbf {\bibinfo {volume} {96}} (\bibinfo {year}
  {2017}{\natexlab{b}}),\ 10.1103/PhysRevA.96.040501}\BibitemShut {NoStop}%
\bibitem [{\citenamefont {Bulgac}\ and\ \citenamefont
  {Efimov}(1976)}]{bulgac:1976_Sov.J.Nucl.Phys.}%
  \BibitemOpen
  \bibfield  {author} {\bibinfo {author} {\bibfnamefont {A}~\bibnamefont
  {Bulgac}}\ and\ \bibinfo {author} {\bibfnamefont {V}~\bibnamefont {Efimov}},\
  }\bibfield  {title} {\enquote {\bibinfo {title} {Spin dependence of the level
  spectrum of three resonantly interacting particles},}\ }\href@noop {}
  {\bibfield  {journal} {\bibinfo  {journal} {Sov. J. Nucl. Phys.}\ }\textbf
  {\bibinfo {volume} {22}},\ \bibinfo {pages} {153} (\bibinfo {year}
  {1976})}\BibitemShut {NoStop}%
\bibitem [{\citenamefont {Kievsky}\ and\ \citenamefont
  {Gattobigio}(2016)}]{kievsky:2016_Few-BodySyst}%
  \BibitemOpen
  \bibfield  {author} {\bibinfo {author} {\bibfnamefont {A.}~\bibnamefont
  {Kievsky}}\ and\ \bibinfo {author} {\bibfnamefont {M.}~\bibnamefont
  {Gattobigio}},\ }\bibfield  {title} {\enquote {\bibinfo {title} {Efimov
  {{Physics}} with $1/2$ {{Spin}}-{{Isospin Fermions}}},}\ }\href {\doibase
  10.1007/s00601-016-1049-5} {\bibfield  {journal} {\bibinfo  {journal}
  {Few-Body Syst}\ }\textbf {\bibinfo {volume} {57}},\ \bibinfo {pages}
  {217--227} (\bibinfo {year} {2016})}\BibitemShut {NoStop}%
\bibitem [{\citenamefont {Kievsky}\ and\ \citenamefont
  {Gattobigio}(2018)}]{kievsky:2018_J.Phys.Conf.Ser.}%
  \BibitemOpen
  \bibfield  {author} {\bibinfo {author} {\bibfnamefont {A.}~\bibnamefont
  {Kievsky}}\ and\ \bibinfo {author} {\bibfnamefont {M.}~\bibnamefont
  {Gattobigio}},\ }\bibfield  {title} {\enquote {\bibinfo {title} {1/2
  spin-isospin fermions close to the unitary limit},}\ }\href {\doibase
  10.1088/1742-6596/981/1/012021} {\bibfield  {journal} {\bibinfo  {journal}
  {Journal of Physics: Conference Series}\ }\textbf {\bibinfo {volume} {981}},\
  \bibinfo {pages} {012021} (\bibinfo {year} {2018})}\BibitemShut {NoStop}%
\bibitem [{\citenamefont {K\"onig}\ \emph {et~al.}(2017)\citenamefont
  {K\"onig}, \citenamefont {Grie\ss{}hammer}, \citenamefont {Hammer},\ and\
  \citenamefont {{van Kolck}}}]{konig:2017_Phys.Rev.Lett.}%
  \BibitemOpen
  \bibfield  {author} {\bibinfo {author} {\bibfnamefont {Sebastian}\
  \bibnamefont {K\"onig}}, \bibinfo {author} {\bibfnamefont {Harald~W.}\
  \bibnamefont {Grie\ss{}hammer}}, \bibinfo {author} {\bibfnamefont {H.-W.}\
  \bibnamefont {Hammer}}, \ and\ \bibinfo {author} {\bibfnamefont
  {U.}~\bibnamefont {{van Kolck}}},\ }\bibfield  {title} {\enquote {\bibinfo
  {title} {Nuclear {{Physics Around}} the {{Unitarity Limit}}},}\ }\href
  {\doibase 10.1103/PhysRevLett.118.202501} {\bibfield  {journal} {\bibinfo
  {journal} {Phys. Rev. Lett.}\ }\textbf {\bibinfo {volume} {118}} (\bibinfo
  {year} {2017}),\ 10.1103/PhysRevLett.118.202501}\BibitemShut {NoStop}%
\bibitem [{\citenamefont {{van Kolck}}(2018)}]{vankolck:2018_J.Phys.Conf.Ser.}%
  \BibitemOpen
  \bibfield  {author} {\bibinfo {author} {\bibfnamefont {U}~\bibnamefont {{van
  Kolck}}},\ }\bibfield  {title} {\enquote {\bibinfo {title} {Nuclear physics
  from an expansion around the unitarity limit},}\ }\href {\doibase
  10.1088/1742-6596/966/1/012014} {\bibfield  {journal} {\bibinfo  {journal}
  {Journal of Physics: Conference Series}\ }\textbf {\bibinfo {volume} {966}},\
  \bibinfo {pages} {012014} (\bibinfo {year} {2018})}\BibitemShut {NoStop}%
\bibitem [{\citenamefont {Hammer}(2018)}]{hammer:2018_Few-BodySyst.}%
  \BibitemOpen
  \bibfield  {author} {\bibinfo {author} {\bibfnamefont {H.-W.}\ \bibnamefont
  {Hammer}},\ }\bibfield  {title} {\enquote {\bibinfo {title} {Nuclei and the
  {{Unitary Limit}}},}\ }\href {\doibase 10.1007/s00601-018-1386-7} {\bibfield
  {journal} {\bibinfo  {journal} {Few-Body Systems}\ }\textbf {\bibinfo
  {volume} {59}} (\bibinfo {year} {2018}),\
  10.1007/s00601-018-1386-7}\BibitemShut {NoStop}%
\bibitem [{\citenamefont {Beane}\ \emph {et~al.}(2001)\citenamefont {Beane},
  \citenamefont {Bedaque}, \citenamefont {Childress}, \citenamefont
  {Kryjevski}, \citenamefont {McGuire},\ and\ \citenamefont {{van
  Kolck}}}]{beane:2001_Phys.Rev.A}%
  \BibitemOpen
  \bibfield  {author} {\bibinfo {author} {\bibfnamefont {S.~R.}\ \bibnamefont
  {Beane}}, \bibinfo {author} {\bibfnamefont {P.~F.}\ \bibnamefont {Bedaque}},
  \bibinfo {author} {\bibfnamefont {L.}~\bibnamefont {Childress}}, \bibinfo
  {author} {\bibfnamefont {A.}~\bibnamefont {Kryjevski}}, \bibinfo {author}
  {\bibfnamefont {J.}~\bibnamefont {McGuire}}, \ and\ \bibinfo {author}
  {\bibfnamefont {U.}~\bibnamefont {{van Kolck}}},\ }\bibfield  {title}
  {\enquote {\bibinfo {title} {Singular potentials and limit cycles},}\ }\href
  {\doibase 10.1103/PhysRevA.64.042103} {\bibfield  {journal} {\bibinfo
  {journal} {Phys. Rev. A}\ }\textbf {\bibinfo {volume} {64}},\ \bibinfo
  {pages} {042103} (\bibinfo {year} {2001})}\BibitemShut {NoStop}%
\bibitem [{\citenamefont {{\'Alvarez-Rodr\'iguez}}\ \emph
  {et~al.}(2016)\citenamefont {{\'Alvarez-Rodr\'iguez}}, \citenamefont
  {Deltuva}, \citenamefont {Gattobigio},\ and\ \citenamefont
  {Kievsky}}]{alvarez-rodriguez:2016_Phys.Rev.A}%
  \BibitemOpen
  \bibfield  {author} {\bibinfo {author} {\bibfnamefont {R.}~\bibnamefont
  {{\'Alvarez-Rodr\'iguez}}}, \bibinfo {author} {\bibfnamefont
  {A.}~\bibnamefont {Deltuva}}, \bibinfo {author} {\bibfnamefont
  {M.}~\bibnamefont {Gattobigio}}, \ and\ \bibinfo {author} {\bibfnamefont
  {A.}~\bibnamefont {Kievsky}},\ }\bibfield  {title} {\enquote {\bibinfo
  {title} {Matching universal behavior with potential models},}\ }\href
  {\doibase 10.1103/PhysRevA.93.062701} {\bibfield  {journal} {\bibinfo
  {journal} {Phys. Rev. A}\ }\textbf {\bibinfo {volume} {93}},\ \bibinfo
  {pages} {062701} (\bibinfo {year} {2016})}\BibitemShut {NoStop}%
\bibitem [{\citenamefont {Kievsky}\ \emph {et~al.}(1997)\citenamefont
  {Kievsky}, \citenamefont {Marcucci}, \citenamefont {Rosati},\ and\
  \citenamefont {Viviani}}]{kievsky:1997_Few-BodySyst}%
  \BibitemOpen
  \bibfield  {author} {\bibinfo {author} {\bibfnamefont {A.}~\bibnamefont
  {Kievsky}}, \bibinfo {author} {\bibfnamefont {L.~E.}\ \bibnamefont
  {Marcucci}}, \bibinfo {author} {\bibfnamefont {S.}~\bibnamefont {Rosati}}, \
  and\ \bibinfo {author} {\bibfnamefont {M.}~\bibnamefont {Viviani}},\
  }\bibfield  {title} {\enquote {\bibinfo {title} {High-{{Precision
  Calculation}} of the {{Triton Ground State Within}} the
  {{Hyperspherical}}-{{Harmonics Method}}},}\ }\href {\doibase
  10.1007/s006010050049} {\bibfield  {journal} {\bibinfo  {journal} {Few-Body
  Syst}\ }\textbf {\bibinfo {volume} {22}},\ \bibinfo {pages} {1--10} (\bibinfo
  {year} {1997})}\BibitemShut {NoStop}%
\bibitem [{\citenamefont {Gattobigio}\ \emph
  {et~al.}(2009{\natexlab{a}})\citenamefont {Gattobigio}, \citenamefont
  {Kievsky}, \citenamefont {Viviani},\ and\ \citenamefont
  {Barletta}}]{gattobigio:2009_Phys.Rev.A}%
  \BibitemOpen
  \bibfield  {author} {\bibinfo {author} {\bibfnamefont {M.}~\bibnamefont
  {Gattobigio}}, \bibinfo {author} {\bibfnamefont {A.}~\bibnamefont {Kievsky}},
  \bibinfo {author} {\bibfnamefont {M.}~\bibnamefont {Viviani}}, \ and\
  \bibinfo {author} {\bibfnamefont {P.}~\bibnamefont {Barletta}},\ }\bibfield
  {title} {\enquote {\bibinfo {title} {Harmonic hyperspherical basis for
  identical particles without permutational symmetry},}\ }\href {\doibase
  10.1103/PhysRevA.79.032513} {\bibfield  {journal} {\bibinfo  {journal} {Phys.
  Rev. A}\ }\textbf {\bibinfo {volume} {79}},\ \bibinfo {pages} {032513}
  (\bibinfo {year} {2009}{\natexlab{a}})}\BibitemShut {NoStop}%
\bibitem [{\citenamefont {Gattobigio}\ \emph
  {et~al.}(2009{\natexlab{b}})\citenamefont {Gattobigio}, \citenamefont
  {Kievsky}, \citenamefont {Viviani},\ and\ \citenamefont
  {Barletta}}]{gattobigio:2009_Few-BodySyst.}%
  \BibitemOpen
  \bibfield  {author} {\bibinfo {author} {\bibfnamefont {M.}~\bibnamefont
  {Gattobigio}}, \bibinfo {author} {\bibfnamefont {A.}~\bibnamefont {Kievsky}},
  \bibinfo {author} {\bibfnamefont {M.}~\bibnamefont {Viviani}}, \ and\
  \bibinfo {author} {\bibfnamefont {P.}~\bibnamefont {Barletta}},\ }\bibfield
  {title} {\enquote {\bibinfo {title} {Non-symmetrized {{Basis Function}} for
  {{Identical Particles}}},}\ }\href {\doibase 10.1007/s00601-009-0045-4}
  {\bibfield  {journal} {\bibinfo  {journal} {Few-Body Syst.}\ }\textbf
  {\bibinfo {volume} {45}},\ \bibinfo {pages} {127--131} (\bibinfo {year}
  {2009}{\natexlab{b}})}\BibitemShut {NoStop}%
\bibitem [{\citenamefont {Gattobigio}\ \emph
  {et~al.}(2011{\natexlab{b}})\citenamefont {Gattobigio}, \citenamefont
  {Kievsky},\ and\ \citenamefont {Viviani}}]{gattobigio:2011_Phys.Rev.C}%
  \BibitemOpen
  \bibfield  {author} {\bibinfo {author} {\bibfnamefont {M.}~\bibnamefont
  {Gattobigio}}, \bibinfo {author} {\bibfnamefont {A.}~\bibnamefont {Kievsky}},
  \ and\ \bibinfo {author} {\bibfnamefont {M.}~\bibnamefont {Viviani}},\
  }\bibfield  {title} {\enquote {\bibinfo {title} {Nonsymmetrized
  hyperspherical harmonic basis for an {{A}}-body system},}\ }\href {\doibase
  10.1103/PhysRevC.83.024001} {\bibfield  {journal} {\bibinfo  {journal} {Phys.
  Rev. C}\ }\textbf {\bibinfo {volume} {83}},\ \bibinfo {pages} {024001}
  (\bibinfo {year} {2011}{\natexlab{b}})}\BibitemShut {NoStop}%
\bibitem [{\citenamefont {Varga}\ and\ \citenamefont
  {Suzuki}(1995)}]{varga:1995_Phys.Rev.C}%
  \BibitemOpen
  \bibfield  {author} {\bibinfo {author} {\bibfnamefont {K.}~\bibnamefont
  {Varga}}\ and\ \bibinfo {author} {\bibfnamefont {Y.}~\bibnamefont {Suzuki}},\
  }\bibfield  {title} {\enquote {\bibinfo {title} {Precise solution of few-body
  problems with the stochastic variational method on a correlated {{Gaussian}}
  basis},}\ }\href {\doibase 10.1103/PhysRevC.52.2885} {\bibfield  {journal}
  {\bibinfo  {journal} {Phys. Rev. C}\ }\textbf {\bibinfo {volume} {52}},\
  \bibinfo {pages} {2885--2905} (\bibinfo {year} {1995})}\BibitemShut {NoStop}%
\bibitem [{\citenamefont {Deltuva}(2013)}]{deltuva:2013_Few-BodySyst}%
  \BibitemOpen
  \bibfield  {author} {\bibinfo {author} {\bibfnamefont {A.}~\bibnamefont
  {Deltuva}},\ }\bibfield  {title} {\enquote {\bibinfo {title} {Properties of
  {{Universal Bosonic Tetramers}}},}\ }\href {\doibase
  10.1007/s00601-012-0313-6} {\bibfield  {journal} {\bibinfo  {journal}
  {Few-Body Syst}\ }\textbf {\bibinfo {volume} {54}},\ \bibinfo {pages}
  {569--577} (\bibinfo {year} {2013})}\BibitemShut {NoStop}%
\bibitem [{\citenamefont {Gattobigio}\ \emph {et~al.}(2019)\citenamefont
  {Gattobigio}, \citenamefont {G\"obel}, \citenamefont {Hammer},\ and\
  \citenamefont {Kievsky}}]{gattobigio:2019_arXiv}%
  \BibitemOpen
  \bibfield  {author} {\bibinfo {author} {\bibfnamefont {M.}~\bibnamefont
  {Gattobigio}}, \bibinfo {author} {\bibfnamefont {M.}~\bibnamefont {G\"obel}},
  \bibinfo {author} {\bibfnamefont {H.-W.}\ \bibnamefont {Hammer}}, \ and\
  \bibinfo {author} {\bibfnamefont {A.}~\bibnamefont {Kievsky}},\ }\bibfield
  {title} {\enquote {\bibinfo {title} {More on the universal equation for
  {{Efimov}} states},}\ }\href {http://arxiv.org/abs/1903.05493} {\bibfield
  {journal} {\bibinfo  {journal} {arXiv}\ } (\bibinfo {year} {2019})},\ \Eprint
  {http://arxiv.org/abs/1903.05493} {arXiv:1903.05493 [cond-mat,
  physics:nucl-th]} \BibitemShut {NoStop}%
\bibitem [{\citenamefont {Gattobigio}\ and\ \citenamefont
  {Kievsky}(2014{\natexlab{a}})}]{gattobigio:2014_Phys.Rev.A}%
  \BibitemOpen
  \bibfield  {author} {\bibinfo {author} {\bibfnamefont {M.}~\bibnamefont
  {Gattobigio}}\ and\ \bibinfo {author} {\bibfnamefont {A.}~\bibnamefont
  {Kievsky}},\ }\bibfield  {title} {\enquote {\bibinfo {title} {Universality
  and scaling in the \${{N}}\$-body sector of {{Efimov}} physics},}\ }\href
  {\doibase 10.1103/PhysRevA.90.012502} {\bibfield  {journal} {\bibinfo
  {journal} {Phys. Rev. A}\ }\textbf {\bibinfo {volume} {90}},\ \bibinfo
  {pages} {012502} (\bibinfo {year} {2014}{\natexlab{a}})}\BibitemShut
  {NoStop}%
\bibitem [{\citenamefont {Gattobigio}\ and\ \citenamefont
  {Kievsky}(2014{\natexlab{b}})}]{gattobigio:2014_JournalofPhysics:ConferenceSeries}%
  \BibitemOpen
  \bibfield  {author} {\bibinfo {author} {\bibfnamefont {M}~\bibnamefont
  {Gattobigio}}\ and\ \bibinfo {author} {\bibfnamefont {A}~\bibnamefont
  {Kievsky}},\ }\bibfield  {title} {\enquote {\bibinfo {title} {Some aspects of
  universality in {{Efimov}} physics},}\ }\href {\doibase
  10.1088/1742-6596/527/1/012002} {\bibfield  {journal} {\bibinfo  {journal}
  {Journal of Physics: Conference Series}\ }\textbf {\bibinfo {volume} {527}},\
  \bibinfo {pages} {012002} (\bibinfo {year} {2014}{\natexlab{b}})}\BibitemShut
  {NoStop}%
\bibitem [{\citenamefont {Ji}\ \emph {et~al.}(2015)\citenamefont {Ji},
  \citenamefont {Braaten}, \citenamefont {Phillips},\ and\ \citenamefont
  {Platter}}]{ji:2015_Phys.Rev.A}%
  \BibitemOpen
  \bibfield  {author} {\bibinfo {author} {\bibfnamefont {Chen}\ \bibnamefont
  {Ji}}, \bibinfo {author} {\bibfnamefont {Eric}\ \bibnamefont {Braaten}},
  \bibinfo {author} {\bibfnamefont {Daniel~R.}\ \bibnamefont {Phillips}}, \
  and\ \bibinfo {author} {\bibfnamefont {Lucas}\ \bibnamefont {Platter}},\
  }\bibfield  {title} {\enquote {\bibinfo {title} {Universal relations for
  range corrections to {{Efimov}} features},}\ }\href {\doibase
  10.1103/PhysRevA.92.030702} {\bibfield  {journal} {\bibinfo  {journal} {Phys.
  Rev. A}\ }\textbf {\bibinfo {volume} {92}} (\bibinfo {year} {2015}),\
  10.1103/PhysRevA.92.030702}\BibitemShut {NoStop}%
\bibitem [{\citenamefont {Pudliner}\ \emph {et~al.}(1995)\citenamefont
  {Pudliner}, \citenamefont {Pandharipande}, \citenamefont {Carlson},\ and\
  \citenamefont {Wiringa}}]{pudliner:1995_Phys.Rev.Lett.}%
  \BibitemOpen
  \bibfield  {author} {\bibinfo {author} {\bibfnamefont {B.~S.}\ \bibnamefont
  {Pudliner}}, \bibinfo {author} {\bibfnamefont {V.~R.}\ \bibnamefont
  {Pandharipande}}, \bibinfo {author} {\bibfnamefont {J.}~\bibnamefont
  {Carlson}}, \ and\ \bibinfo {author} {\bibfnamefont {R.~B.}\ \bibnamefont
  {Wiringa}},\ }\bibfield  {title} {\enquote {\bibinfo {title} {Quantum {{Monte
  Carlo Calculations}} of {{A}} {$\leq$} 6 {{Nuclei}}},}\ }\href {\doibase
  10.1103/PhysRevLett.74.4396} {\bibfield  {journal} {\bibinfo  {journal}
  {Phys. Rev. Lett.}\ }\textbf {\bibinfo {volume} {74}},\ \bibinfo {pages}
  {4396--4399} (\bibinfo {year} {1995})}\BibitemShut {NoStop}%
\bibitem [{\citenamefont {Gattobigio}\ \emph {et~al.}(2012)\citenamefont
  {Gattobigio}, \citenamefont {Kievsky},\ and\ \citenamefont
  {Viviani}}]{gattobigio:2012_Phys.Rev.A}%
  \BibitemOpen
  \bibfield  {author} {\bibinfo {author} {\bibfnamefont {M.}~\bibnamefont
  {Gattobigio}}, \bibinfo {author} {\bibfnamefont {A.}~\bibnamefont {Kievsky}},
  \ and\ \bibinfo {author} {\bibfnamefont {M.}~\bibnamefont {Viviani}},\
  }\bibfield  {title} {\enquote {\bibinfo {title} {Energy spectra of small
  bosonic clusters having a large two-body scattering length},}\ }\href
  {\doibase 10.1103/PhysRevA.86.042513} {\bibfield  {journal} {\bibinfo
  {journal} {Phys. Rev. A}\ }\textbf {\bibinfo {volume} {86}},\ \bibinfo
  {pages} {042513} (\bibinfo {year} {2012})}\BibitemShut {NoStop}%
\bibitem [{\citenamefont {Wiringa}\ \emph {et~al.}(1984)\citenamefont
  {Wiringa}, \citenamefont {Smith},\ and\ \citenamefont
  {Ainsworth}}]{wiringa:1984_Phys.Rev.C}%
  \BibitemOpen
  \bibfield  {author} {\bibinfo {author} {\bibfnamefont {R.~B.}\ \bibnamefont
  {Wiringa}}, \bibinfo {author} {\bibfnamefont {R.~A.}\ \bibnamefont {Smith}},
  \ and\ \bibinfo {author} {\bibfnamefont {T.~L.}\ \bibnamefont {Ainsworth}},\
  }\bibfield  {title} {\enquote {\bibinfo {title} {Nucleon-nucleon potentials
  with and without {{$\Delta$}} ( 1232 ) degrees of freedom},}\ }\href
  {\doibase 10.1103/PhysRevC.29.1207} {\bibfield  {journal} {\bibinfo
  {journal} {Phys. Rev. C}\ }\textbf {\bibinfo {volume} {29}},\ \bibinfo
  {pages} {1207--1221} (\bibinfo {year} {1984})}\BibitemShut {NoStop}%
\bibitem [{\citenamefont {Lu}\ \emph {et~al.}(2018)\citenamefont {Lu},
  \citenamefont {Li}, \citenamefont {Elhatisari}, \citenamefont {Lee},
  \citenamefont {Epelbaum},\ and\ \citenamefont {Mei\ss{}ner}}]{lu:2018_arXiv}%
  \BibitemOpen
  \bibfield  {author} {\bibinfo {author} {\bibfnamefont {Bing-Nan}\
  \bibnamefont {Lu}}, \bibinfo {author} {\bibfnamefont {Ning}\ \bibnamefont
  {Li}}, \bibinfo {author} {\bibfnamefont {Serdar}\ \bibnamefont {Elhatisari}},
  \bibinfo {author} {\bibfnamefont {Dean}\ \bibnamefont {Lee}}, \bibinfo
  {author} {\bibfnamefont {Evgeny}\ \bibnamefont {Epelbaum}}, \ and\ \bibinfo
  {author} {\bibfnamefont {Ulf-G.}\ \bibnamefont {Mei\ss{}ner}},\ }\bibfield
  {title} {\enquote {\bibinfo {title} {Essential elements for nuclear
  binding},}\ }\href {http://arxiv.org/abs/1812.10928} {\bibfield  {journal}
  {\bibinfo  {journal} {arXiv}\ } (\bibinfo {year} {2018})},\ \Eprint
  {http://arxiv.org/abs/1812.10928} {arXiv:1812.10928 [hep-lat,
  physics:nucl-th]} \BibitemShut {NoStop}%
\end{thebibliography}
\end{document}